\documentclass[a4paper,10pt]{article}

 \usepackage[cp1251]{inputenc}
 \usepackage{cite}

 \usepackage{graphicx}

\usepackage{cite,amsmath,amsfonts,amsthm,fullpage}
\usepackage{youngtab}
\newcommand{\Pf}{\mathop\mathrm{Pf}\nolimits}
\newcommand{\sgn}{\mathop\mathrm{sgn}\nolimits}

\newcommand{\bpow}{\mathbf{p}}

\newcommand{\f}{\textsc{f}}
\newcommand{\e}{\textsc{e}}
\newcommand{\V}{\textsc{v}}

\newcommand{\hc}{\textsc{h}}
\newcommand{\mc}{\textsc{m}}

\newcommand{\E}{\mathbb{E}}

\newcommand{\Lz}{\textbf{L}_Z}

\newcommand{\Lu}{\textbf{L}_U}

\def\t{\texttt{t}}
\def\q{\texttt{q}}

\theoremstyle{plain}

\newtheorem{Lemma}{Lemma}
\newtheorem{Proposition}{Proposition}
\newtheorem{Corollary}{Corollary}
\newtheorem{Remark}{Remark}
\newtheorem{Example}{Example}

\theoremstyle{remark}

\def\l{\langle}
\def\r{\rangle}
\def\g{\Gamma}

\def\bp{\begin{Proposition}}
\def\ep{\end{Proposition}}
\def\bc{\begin{Corollary}}
\def\ec{\end{Corollary}}
\def\bl{\begin{Lemma}}
\def\el{\end{Lemma}}
\def\be{\begin{equation}}
\def\ee{\end{equation}}
\def\br{\begin{Remark}\rm\small}
\def\er{\end{Remark}}
\def\brs{\begin{remarks}.\\ \rm\
\begin{enumerate}}
\def\ers{\end{enumerate}\end{remarks}}
\def\bea{\begin{eqnarray}}
\def\eea{\end{eqnarray}}

\def\tr{\mathrm {tr}}
\def\det{\mathrm {det}}

\def\sgn{\mathrm {sgn}}

\def\&{&{\hskip -20pt}}

\usepackage[usenames,dvipsnames]{color}
\usepackage{ulem}

\begin{document}

\author{Sergey M. Natanzon\thanks{National Research University Higher School of Economics, Moscow, Russia; 
Institute for Theoretical and Experimental Physics, Moscow, Russia;
email: natanzons@mail.ru} \and Aleksandr Yu.
Orlov\thanks{P.P. Shirshov Institute of Oceanology, Russian Academy of Sciences,
Nahimovskii Prospekt 36,
Moscow 117997, Russia, and National Research University Higher School of Economics,
International Laboratory of Representation
Theory and Mathematical Physics,
20 Myasnitskaya Ulitsa, Moscow 101000, Russia, email: orlovs@ocean.ru
} }
\title{Integrals of tau functions}

\maketitle

\begin{abstract}

\noindent 

We consider integrals of tau functions of Zakharov-Shabat systems whose
higher times are related to the eigenvalues of products of random matrices.
Apart of random matrices there is the set of $n$ pairs of given matrices 
which play the role of parameters. In terms of these matrices we introduce the 
notions of words, dressed words and dual words, these notions are related to
the graphs and dual graphs drawn on a Riemann surface $\Sigma$. 
 The integrals of tau functions over independent ensembles of random matrices 
 can be computed in form of series over partitions of products 
of the Schur polynomials with a multiplier which depends on the Euler characteristic
of the Riemann surface. This form allows to compare the integrals of tau functions
with correlation functions of certain quantum models. 
We present a tau function whose integral is equal to the correlation function
of the Wilson loops of the two-dimensional Yang-Mills model on $\Sigma$.

\end{abstract}

\bigskip

\textbf{Key words:} Zakharov-Shabat sysytems,  random matrices, tau functions, BKP hierarchy,
  Schur polynomials, hypergeometric functions,
 random partitions, 2D YM

\textbf{2010 Mathematic Subject Classification:} 05A15, 14N10, 17B80, 35Q51, 35Q53, 35Q55, 37K20, 37K30,

\qquad\qquad\qquad\qquad Article is dedicated to the 80th anniversary of Vladimir E. Zakharov

\section{Introduction \label{Introduction}}

The results of the work are presented in Section \ref{Integrals-of-tau-functions}. We show
that integrals of tau functions are also integrable in the sense of \cite{Kazakov-SolvMM}
and may be related to quantum models as 2-dimensional Yang-Mills theory on orientable \cite{Migdal},
\cite{Rusakov},\cite{Witten} and non-orientable \cite{Witten} surfaces.

This work appears as a result of the cross study of three topics: integrable hierarchies
\cite{ZakharovShabat},\cite{NovikovManakovZakharov},\cite{Mikhailov},
\cite{Sato},\cite{DJKM},\cite{JM},\cite{UT},\cite{Takasaki-Schur},
\cite{Takasaki2018},\cite{KvdLbispec},\cite{OS-2000},\cite{OST-I},\cite{OST-II}, matrix models \cite{Kazakov},
\cite{GMMMO},\cite{Morozov-MM-review},\cite{Kazakov-SolvMM},\cite{ZinnJustin},\cite{O-review2002},
\cite{HO-2006} products of random matrices
\cite{Ak1},\cite{Ak2},\cite{AkStrahov},\cite{Alfano}
and Hurwitz numbers in case of orientable surfaces \cite{Dijkgraaf},\cite{Goulden-Jackson-2008},
\cite{Okounkov-2000},\cite{Okounkov-Pand-2006},
\cite{MM1}-\cite{MM5},\cite{AMMN-2014},\cite{HO-2014},\cite{ChekhovAmbjorn} and also of 
nonorientable ones \cite{AN1}-\cite{AN},
\cite{Carrell},
\cite{NO-2014},\cite{NO-LMP}.

The necessary preliminary  data are presented in this section below.

\subsection{Preliminaries}

\paragraph{Zakharov-Shabat systems.}  Let us remind that Zakharov-Shabat $2+1$ integrable systems are constructed 
with the help of '$L-A$ pairs' (or, better to say, Zakharov-Shabat (ZS) pairs) which are differential operators with 
matrix valued coefficients. Zakharov-Shabat operators can be obtained with the help of the dressing procedure
of vacuum differential operators \cite{ZakharovShabat}, \cite{NovikovManakovZakharov}. 
It is suitable to choose the following set of vacuum ZS operators
\[
 \frac{\partial}{\partial p_m^{(i)}}-e_i \left(\frac{\partial}{\partial p_1^{(i)}}\right)^m,\quad m=1,2,\dots
\]
where $e_i$ ($i=1,\dots,\textsc{d}$) is $\textsc{d}\times \textsc{d}$ matrices with all entries vanish except the entry $i,i$ which is equal to 1.
In this case, the  sets of of $\bpow^{i}=(p_1^{(i)},p_2^{(i)},p_3^{(i)},\dots),\,i=1,\dots,\textsc{d}$ are called higher 
times of the (multicomponent) KP hierarchy (the word "multicomponent" is related to the label $i$). 
The Kadomtsev-Petviashvili (KP) equation 
$$
12 u_{p_1p_3}-u_{p_1p_1p_1p_1}-12 u_{p_2,p_2}  -3(u^2)_{p_1p_1}=0
$$
is obtained with the help of the choice of the pair with $m=2$, $m=3$ and $\textsc{d} =1$.
In the setting of applications to the study of the wave on the water surface, $p_1$ and $p_2$ plays the 
role of space $x$ and $y$ variables in the horizontal plane
while $p_3$ plays the role of the time. All other parameters $p_m,\, m>3$ play the role of the group times
of the hidden symmetries which provide the integrability of the KP equation. This hierarchy of compatible
symmetry flows is called KP hierarchy. The hierarchy of the compatible flows with respect to the parameters
(the higher times) $p_m^{(i)},\,i=1,\dots,\textsc{d},\, m>0$ is called the multicomponent KP hierarchy.
The role of sets of higher times in the theory of integrable systems is very important.

To write down the analogue
 of the Kadomtsev-Petviashvili equation in case $\textsc{d}>1$ 
one should choose any three independent variables among six ones $p^{i}_m,\, i=1,2,\, m=1,2,3$ and in general 
 is rather specious. However, among this set one can pick up the subset of the KP equations numbered by $i$:
 \be\label{KP-N-i}
12 u^{(i)}_{p^{(i)}_1p^{(i)}_3}-u^{(i)}_{p^{(i)}_1p^{(i)}_1p^{(i)}_1p^{(i)}_1}
-12 u^{(i)}_{p^{(i)}_2,p^{(i)}_2}  -3\left(\left(u^{(i)} \right)^2\right)_{p^{(i)}_1p^{(i)}_1}=0,\quad i=1,\dots,\textsc{d}
 \ee
 where $u^{(i)}_{p^{(j)}_1p^{(j)}_1}=u^{(j)}_{p^{(i)}_1p^{(i)}_1}$.

 In case $\textsc{d}=2$, there are two ways to get equations in a relatively compact form. 
 The first way to present two-component KP hierarchy as the hierarchy of the two-dimensional Toda lattice (TL) 
 hierarchy (see \cite{UT} for details).
The second way impose certain symmetry
condition on the matrix coefficients. The example which is related to our paper is the reduction
\cite{Kakei-2} which may be called orthogonal reduction where instead of two sets of independent variables 
$\bpow^{(i)}, \, i=1,2$ we get the single one 
denoted by $\bpow$ as in the one-component case, and in this case one gets the following system
\be\label{DKP}
12 u_{p_3p_1}- u_{p_1p_1p_1p_1}-12 u_{p_2p_2}-3 \left(u^2\right)_{p_1p_1}  
 +24(v{\tilde v})_{p_1p_1} =0
\ee
\[
6v_{p_3}+3uv_{p_1} +  v_{p_1p_1p_1} + 6 v_{p_1p_2} +6 v\partial^{-1}_{p_1} u_{p_2}  = 0
\]
\[
6{\tilde v}_{p_3}+3u{\tilde v}_{p_1} +  {\tilde v}_{p_1p_1p_1} 
+ 6 {\tilde v}_{p_1p_2} + 6 {\tilde v}\partial^{-1}_{p_1} u_{p_2}  = 0
\]
At first this system of equations were obtained by Hirota method in \cite{HirotaOhta}. The Zakharov-Shabat representation
 Later it was also recognized (J. van de Leur) as an example of the so-called
DKP hierarchy, namely, the KP equation based on the root system D \cite{JM}, \cite{KvdLbispec}.

\paragraph{Tau functions of Kyoto school.} 
A fruitful approach to a significant number of tasks related to
applications of the theory of solitons in mathematics and modern mathematical physics is the use of
the so-called tau function of the multicomponent KP
hierarchy introduced and developed by Sato school in Kyoto \cite{Sato}, \cite{JM}.
It was noticed in the celebrated works by Zakharov and  by Zakharov and Shabat that soliton
solutions of integrable models can be expressed in form of the logarithmic derivative of certain determinants.
This determinant is an example of the tau function. For $N$-soliton solution this is the determinant of a finite
(basically $N\times N$) matrix, while in general case it is the determinant of an infinite size matrix.

The appearance of determinants invites to present it as a Slater type determinant which describes fermionic
systems. Indeed \cite{JM}, tau functions can be presented as certain correlation functions for free fermion system.

In  terms of tau function the solution of the system (\ref{KP-N-i}) can be written as
$$
u^{(i)}=2\left( \log\tau \right)_{p^{(i)}_1p^{(i)}_1}
$$
In $\textsc{d}=2$ case with the orthogonal reduction $u,v,{\bar v}$ of (\ref{DKP}) are also expressed in terms of the DKP 
tau function, in particular, $u=2\left( \log\tau^{2KP} \right)_{p_1p_1}=4\left( \log\tau^{DKP} \right)_{p_1p_1}$.

\paragraph{Tau function as series in the Schur functions.} In \cite{Sato} it was shown that tau functions
can be presented in form of series in the Schur polynomials (also known as the Schur functions), see \cite{Mac}.
Actually, this is the form to write down the tau function in form of the Taylor series in higher times,
where instead of monomials one uses the basis of the Schur polynomials.
Explicit expressions related to the Schur functions see in the next paragraph.
These polynomials are labeled by multiindices, written as \textit{partitions} 
$\lambda$, $\lambda = (\lambda_1,\dots,\lambda_l)$,
where $\lambda_1\ge \cdots \ge \lambda_l\ge 0$ are nonnegative integers called \textit{parts} of the partition.
The Schur polynomials form the basis in the space of polynomials in many variables.
Tau functions which we will need are of two types. The first type is the tau functions of the 
$\textsc{d}$-component KP \cite{Sato},\cite{JM}.
It may be defined with the help of a given (in general, of an infinite size, however in this case
not far enough from the identity
infinite matrix) matrix $\textbf{A}$, which
consists of $\textsc{d}$ (in general, infinite size) blocks, each block is indexed by $(i,j),\,i,j=1,\dots,\textsc{d}$.
The matrix $\textbf{A}$ plays the role of the initial
data for the equations of the $\textsc{d}$-component KP 
(see the detailed analisis of $\textsc{d}=1$ case in \cite{UT}.\cite{Takasaki-Schur} and the study
of the related fermionic correlators in \cite{TI}-\cite{TII}). 
The tau function can be written as follows:
\be\label{tau-F-KP'}
\tau^\textbf{A}_{n_1,\dots,n_\textsc{d}}(\bpow^{1},\dots,\bpow^{\textsc{d}})=
\sum_{\lambda^1,\dots,\lambda^\textsc{d}} \textbf{A}_{\lambda^1,\dots,\lambda^\textsc{d}}(n_1,\dots,n_\textsc{d})
\prod_{i=1}^\textsc{d} s_{\lambda^i}(\bpow^i)
\ee
where the notations are as follows. 
Each Schur function $s_{\lambda^i}$ is a polynomial of the set of higher times indexed by $i$, $\bpow^i$.
Each prefactor $\textbf{A}_{\lambda^1,\dots,\lambda^\textsc{d}}(n_1,\dots,n_\textsc{d})$ is the determinant of the
submatrix of $\textbf{A}$,
where the choice of the submatrix is defined by partions $\lambda^1,\dots,\lambda^\textsc{d}$ and the set of
integers $n_1,\dots,n_\textsc{d}$ which plays the role of discrete higher times additional to the sets $\bpow^1,
\dots,\bpow^\textsc{d}$.

For the sake of simplicity, we will restrict ourselves with the case where 
 $n^1=\cdots = n^\textsc{d}=0$ and omit the dependence on these numbers. 
 In this case the entries of each $i,j$ block of $\textbf{A}$ we 
 denote $\textbf{A}^{i,j}_{a,b}$ where $a,b>0$.
it is natural to use the Frobenius coordinates of the partitions $\lambda^i=(\alpha^i|\beta^i)$, see \cite{Mac},
namely sets of arm coordinated $\alpha^i=(\alpha^i_1,\dots,\alpha^i_{\kappa^{(i)}})$, where
$\alpha^i_1>\dots >\alpha^i_{\kappa^{(i)}}\ge 0$, and leg coordinated $\beta^i=(\beta^i_1,\dots,\beta^i_{\kappa^{(i)}})$, where
$\beta^i_1>\dots >\beta^i_{\kappa^{(i)}}\ge 0$ and $\kappa^{(i)}$ is the number of nodes in the main diagonal of the Young diagram
of $\lambda^i$. The submatrix whose determinant is
$\textbf{A}_{\lambda^1,\dots,\lambda^\textsc{d}}$ is $K\times K$ matrix where the $K=\sum_{i=1}^\textsc{d} \kappa^{(i)}$
with the set of  entries $\{\alpha^{(i)}_{l^{(i)}},\,\beta^{(j)}_{m^{(j)}}\}$ where
$l^{(i)}=1,\dots,\kappa^{(i)}$ and  $m^{(i)}=1,\dots,\kappa^{(j)}$ in each of $i,j$ block of $\textbf{A}$:
\be\label{A-prefactor}
\textbf{A}_{\lambda^1,\dots,\lambda^\textsc{d}}(0,\dots,0)=
\textbf{A}_{\lambda^1,\dots,\lambda^\textsc{d}}
=
\det 
\left(\textbf{A}^{i,j}_{\alpha^{(i)}_{l^{(i)}},\beta^{(j)}_{m^{(j)}} }
\right)_{i,j=
1,\dots,\textsc{d},\,l^{(i)}=1,\dots,\kappa^{(i)},m^{(j)}=1,\dots,\kappa^{(j)} }
\ee

$\quad$

The second type of tau functions generalizes tau functions of system (\ref{DKP}) and is related
to the orthogonal reduction of $2\textsc{d}$-component KP hierarchy (see \cite{LO-LMP} for $\textsc{d}=1$ case).
The square root of $2\textsc{d}$-component tau function is called tau function of the $\textsc{d}$-component BKP hierarchy
introduced in \cite{KvdLbispec}.

$\textsc{d}$-component BKP tau function may be defined by (in general, infinite size) skew-symmetric matrix 
$\textbf{B}$ which consists of (in general, infinite size) blocks $(i,j),\, i,j=\pm 1,\dots,\pm\textsc{d}$.
It may be written as
\[
\tau_{n_{1},\dots,n_{2\textsc{d}}}^\textbf{B}(\bpow^{1},\dots,\bpow^{\textsc{d}})=
1+\sum_{\lambda^1,\dots,\lambda^\textsc{d}} \textbf{B}_{\lambda^1,\dots,\lambda^\textsc{d}}(n_{1},\dots,n_{2\textsc{d}})
\prod_{i=1}^\textsc{d} s_{\lambda^i}(\bpow^i)
\]
where $\textbf{B}_{\lambda^1,\dots,\lambda^\textsc{d}}$ is the pfaffian of the submatrix of $\textbf{B}$,
the entries of this submatrix are defined with the help of the whole set of partitions 
$\lambda^i\,i=1,\dots,\textsc{d}$. However, in this case we restrict ourselves by the case $\textsc{d}=1$
and rather simple BKP tau function we call BKP hypergeometric tau function \cite{OST-I}, see below.

\br\label{convergency}
The series over partition should be considered as formal ones. However, typically (say, if $\textbf{A}$ is 
close enough to the identity matrix, or in other words $\textsc{A}$ belongs to a properly defined 
$\in\mathbb{GL}_\infty$ group) there are open 
domains of convergency in the space of higher times $\bpow^i,\,i=1,\dots,2\textsc{d}$. Notice that
\[
 s_{\lambda^i}(\bpow^i)=0\qquad {\rm if}\qquad p^{(i)}_m =\pm\sum_{j=1}^k x_j^m
\]
for $\ell(\lambda)>k$ in case "+" factor and for  $\lambda^{(i)}_1>k$ in case "-" factor.
For solutions in form of formal series KP and BKP equations have the sense of recursion equation. 
\er

\paragraph{Schur functions.}
In what follows $N$ is a fixed number.
In what follows we need polynomials in many variables $\bpow=(p_1,p_2,\dots )$
called Schur functions labelled with partitions \cite{Mac}. To define these polynomials let us consider
\be\label{tttV}
\texttt{V}(x,\bpow):=\sum_{m>0} \frac 1m x^m p_m
\ee
and introduce the so-called elementary Schur functions
$ s_{(n)} $, labelled by $ (n) $, that is by partitions with a single part $ \lambda_1=n $ as follows:
$$
e^{\texttt{V}(x,\bpow)}=\sum_{n\ge 0} x^n s_{(n)}(\bpow)
$$
In particular, $s_{(0)}(\bpow)=1$,  $s_{(1)}(\bpow)=p_1$, $s_{(2)}(\bpow)=\frac12 (p_1^2+ p_2)$.

Schur function $s_\lambda$ labelled by a given partition $\lambda=(\lambda_1,\dots ,\lambda_N)$ is defined
in terms of the elementary ones by
\be\label{SchurFunction}
s_\lambda(\bpow) = \det \left( s_{(\lambda_i-i+j)}(\bpow) \right)
\ee
where $i,j=1,\dots,N$, $N$ is any number exceeding the length $\ell(\lambda)$ (we recall that the length
$\ell(\lambda)$ is the number of the non-vanishing parts of 
$\lambda=(\lambda_1,\dots,\lambda_k\ge 0),\,k=\ell(\lambda)$.
Let us denote the Young diagram obtained from the Young diagram $\lambda$ by reflection with respect to
the main diagonal $\lambda^{t}$. Then there is the following property which can be obtained from the definition
\[
s_{\lambda^t}(-\bpow)=(-1)^{|\lambda|}s_\lambda(\bpow),\quad |\lambda|=\sum_{i}\lambda_i
\]
or, the same in the Frobenius coordinated for partitions:
\be\label{reflection}
s_{(\alpha|\beta)}(-\bpow)=(-1)^{|\lambda|}s_{(\beta|\alpha))}(\bpow)
\ee
and $|\lambda|=|\alpha|+|\beta|+\kappa$, where $\kappa$ is the number of nodes in the main diagonal of the Young
diagram of $\lambda$. (We recall that the Frobenius coordinated are the lengths of arms and legs of the hooks
whoose corners are the nodes of the main diagonal of the Young diagram of $\lambda$, see \cite{Mac}).

We shall write the Schur function also as the function of matrix argument which we write as a capital letter
 say $X$ having in mind that  
 \be\label{convention}
 s_\lambda(X):=s_\lambda(\bpow(X))
 \ee
 (see (\ref{p(X)})) where we use the convention
 \be\label{p(X)}
 \bpow(X)=\left(p_1(X),p_2(X),\dots  \right)\quad {\rm where} \quad p_n(X)=\tr X^n
 \ee
  If $x_1,\dots,x_N$ are
 the eigenvalues of the $N\times N$ matrix $X\in\mathbb{GL}_N(\mathbb{C})$, then $s_\lambda(X)$ is the 
 symmetric homogenious polynomial 
 in eigenvalues and can be written as
 \be\label{Schur-in-X}
 s_\lambda(X)= \frac{\det \left(x_j^{N+\lambda_i-i}  \right)}
 {\det \left(x_j^{N-i} \right)}
 \ee
 In this formula $i,j=1,\dots,N$. It is implied that the length of $\lambda$ does not exceed $N$,
 otherwise the Schur function vanishes. It can be derived from (\ref{SchurFunction}) where we substitue
 $\bpow=\bpow(X)$.
 
 The formula known as Cauchy-Littlewood relation \cite{Mac}  is very useful  
\be\label{KP-vac}
e^{\sum_{m>0} \frac 1m p_m{\tilde p}_m}=\sum_{\lambda} s_\lambda(\bpow)s_\lambda({\tilde\bpow})=
1+s_{(1)}(\bpow)s_{(1)}({\tilde\bpow})+\cdots
\ee
In case ${\tilde\bpow}=\bpow(X)$ this relation takes the form 
 \be\label{C-L}
 e^{N \tr \texttt{V}(X,\bpow)} =\sum_\lambda s_\lambda(X)s_\lambda(N\bpow)
 \ee
 where the sum ranges over all partitions whose length (the number of non-vanishing parts) does not exceed $N$,
 and $N\bpow:=\left( Np_1, Np_2, Np_3,\dots \right)$.

 Another equality which is just the Taylor expansion of the exponential is of use:
\be\label{Taylor-for-e^V}
e^{N\tr \texttt{V}(X,\bpow)} = \sum_\Delta \frac{1}{z_\Delta} N^{\ell(\Delta)} \bpow_\Delta(X)\bpow_\Delta
\ee
where $z_\Delta$ is defined by 
\be\label{z_Delta}
 z_\Delta =\prod_{i\ge 1} i^{m_i}m_i!
\ee
where $m_i$ is the number of times the number $i$ occurs in the partition $\Delta$.
The sum (\ref{Taylor-for-e^V}) ranges over all partitions $\Delta=(\Delta_1,\Delta_2,\dots, \Delta_k)$, $\Delta_k >0$,
$k=0,1,2,\dots$. As usual,
$\ell(\Delta)$ denotes the {\it length} of the partition $\Delta$, i.e. the number of the non-vanishing
parts of $\Delta$.

\paragraph{Characteristic map relation.}  There exists the wonderful relation between the characters of the linear 
and of the symmetric groups:
\be\label{Schur-char-map}
s_\lambda\left(\bpow(X)\right)=
\frac{\operatorname{dim}_S\lambda}{d!}\,\sum_{\Delta\atop |\Delta|=|\lambda| } 
\varphi_\lambda(\Delta)\bpow_{\Delta}(X),\quad d=|\lambda|
\ee
or, the same,
\be\label{Schur-char-map'-opposite}
\bpow(X)=
\frac{\operatorname{dim}_S\lambda}{d!}\,\sum_{\lambda\atop |\Delta|=|\lambda| } 
\varphi_\lambda(\Delta)s_\lambda(X)
\ee
where ${\rm dim}_S\lambda$ is the dimension of the irreducible representation of the symmetric group 
$S_d$ labeled by $\lambda$, and  $X\in \mathbb{GL}_N(\mathbb{C})$, see for instance \cite{Mac}. This relation does depend on $N$ and is still 
correct if we replace
the set $\bpow(X)$ by any set of variables $\bpow=(p_1,p_2,\dots)$. Here
\[
 \varphi_\lambda(\Delta) = \frac{1}{z_\Delta}\frac{|\lambda|!}{{\rm \dim}_S\lambda}\chi_\lambda(\Delta)
\]
where $|\lambda|$ is the weight of the partition $\lambda$ (i.e. 
the number of nodes of the Young diagram of $\lambda$), $\chi_\lambda(\Delta)$ is the character of
the irreducible representation of the symmetric group $S_d$ labeled by $\lambda$ evaluated of the element of the symmetric group
$S_d$ of the cycle class $\Delta$ (i.e. this element can be presented as a product of
nonintersecting cycles whose lengths are $\Delta_1,\Delta_2,\dots$, $\Delta=(\Delta_1,\Delta_2,\dots)$ , 
$\Delta_1\ge \Delta_2 \ge \cdots $).

\paragraph{Content product.} For a given number $x$ and a given Young diagram $\lambda$ the content 
product is defined as the product
\be\label{Poch-YD}
(x)_\lambda :=\prod_{(i,j)\in \lambda} (x+j-i) 
\ee
The number $j-i$, which is the distance of the node with coordinates $(i,j)$ to the main diagonal of the Young
diagram $\lambda$ is called the {\it content} of the node. For one-row $\lambda$, the content product is the 
Pochhammer symbol $(a)_{\lambda_1}$. For a given function of one variable $r$, we define the {\it generalized
content product} (the generalized Pochhammer symbol) as
\be\label{content-product}
r_\lambda(x)= \prod_{(i,j)\in\lambda} r(x+j-i)
\ee

The content product plays an important role in the representation theory of the symmetric groups.
It was used in \cite{OS-2000} to define certain family of tau functions which we called hypergeometric
tau functions.
 
\paragraph{Content products in terms of the Schur functions evaluated at special points.} The example of 
the generalized content product may be 
constructed purely in terms of the Schur functions:
if we choose 
\be\label{Example:q,t}
r(x)=\prod_{i} \left(\frac{1-{  \t  }_i{\q}_i^x}{1-  {\q}_i^x}  \right)^{\texttt{d}_i}
\ee
where $ {  \t  }_i,{\q}_i,\texttt{d}_i$ are parameters, we obtain
\be\label{r-rational}
r_\lambda(x)=\prod_i 
\left( \frac{s_\lambda(\bpow ({  \t  }_i,{\q}_i))} {s_\lambda(\bpow (0,{\q}_i))} 
\right)^{\texttt{d}_i}
\ee
One can degenerate (\ref{Example:q,t}) to the rational function and obtain
\be\label{r-lambda-rational}
r_\lambda(x)= \frac{\prod_{i=1}^p ({\texttt{a}}_i)_\lambda}{\prod_{i=1}^q (\texttt{b}_i)_\lambda}=
\prod_{i=1}^p \frac{s_\lambda(\bpow(\texttt{a}_i))}{s_\lambda(\bpow_\infty)}
\prod_{i=1}^q \frac{s_\lambda(\bpow_\infty)}{s_\lambda(\bpow(\texttt{b}_i))}
\ee
Above we used the following special notations:
\be\label{special-p}
\bpow_\infty=(1,0,0,\dots),\quad \bpow(\texttt{a})=(\texttt{a},\texttt{a},\texttt{a},\dots),\quad 
p_m({  \t  },{\q})=\frac{1-{  \t  }^m}{1-{\q}^m}
\ee
Actually, any reasonable content product can be interpolated by expressions (\ref{r-lambda-rational}).
\br\label{p=power-sums-f-N-terms} Notice, that $\bpow(\q^N,\q)=1+\q+\dots +\q^{N-1}$ which allows
to interpret it as $\bpow(X)$ where $1,\dots,\q^{N-1}$ are eigenvalues of $X$.
\er

Let us also write down the known formula \cite{Mac}
\be\label{Schur-dim}
s_\lambda(\bpow(a))=s_\lambda(\bpow_\infty)\prod_{(i,j)\in\lambda}\left(a+j-i \right),\quad
 s_\lambda(\bpow_\infty)=\frac{{\rm dim}_S\lambda}{d!}=\frac{\prod_{i<j}^N(h_i-h_j)}{\prod_{i=1}^N h_i!}
\ee
\be\label{Schur-p(q,t)}
 s_\lambda(\bpow({  \t  },{\q}))=s_\lambda(\bpow(0,{\q}))\prod_{(i,j)\in\lambda}
 \left(1-{  \t  }{\q}^{j-i}  \right),\quad 
 s_\lambda(\bpow(0,{\q}))={\q}^{n(\lambda)}
 \frac{\prod_{i<j}^N(1-{\q}^{h_i-h_j})}{\prod_{i=1}^N ({\q};{\q})_{h_i}}
\ee
where $n(\lambda):=\sum_{i=1}^{\ell(\lambda)}(i-1)\lambda_i$
\be\label{t-Pochhammer}
(a;{\q})_{n}:=(1-a)(1-a{\q})\cdots (1-a{\q}^{n-1})
\ee
and
where $d=|\lambda|:=\sum_{i=1}^N \lambda_i$ is the weight of $\lambda$, and $h_i=\lambda_i-i+N,\,i=1,\dots,N$ 
are called the {\it shifted parts} of the partition $\lambda$, we imply $N\ge\ell(\lambda)$. 
\br\label{saturation}
Formulae (\ref{Schur-dim}) and (\ref{Schur-p(q,t)}) possess the saturation property: they does not depend on the 
choise of $N$ if $N$ is large enough.
\er

Let us write down useful relations which follow from (\ref{Schur-char-map}) and (\ref{special-p}):
\be
s_\lambda(\bpow(a))=s_\lambda(\bpow_\infty)\sum_{k=0}^{d}\phi_k(\lambda) a^k
\ee
where
\be
\phi_k(\lambda):=\sum_{\Delta} \varphi_{\Delta}(\lambda)
\ee
and
\be\label{q-t-Schur-char-map}
s_\lambda(\bpow(\texttt(q),{\q}))=s_\lambda(\bpow_\infty)
\sum_{\Delta\atop |\Delta|=|\lambda|} \varphi_\lambda(\Delta) 
\prod_{i=1}^{\ell(\Delta)}
\left(\frac{1-{  \t  }^{\Delta_i}}{1-{\q}^{\Delta_i}}\right)   
\ee

\subsection{Complex Ginibre ensembles, ensembles of unitary matrices and mixed ensembles \label{random-complex}}

Let us consider integrals over  $N\times N$ complex matrix $Z$ where the measure is defined as
\be\label{CGE-measure}
d\mu(Z)=c_N \prod_{i,j=1}^N d\Re Z_{ij}d\Im Z_{ij}\text{e}^{-N|(Z_\alpha)_{ij}|^2}
\ee
where the integration range is $\mathbb{C}^{N^2}$ and where $c_N$
is the normalization constant defined via $\int d \mu(Z)=1$, thus 
we treat this measure as the probability measure. 
In studies of Quantum Chaos and in studies for Information Transmission Problems,
the set of complex matricies with such measure is called the complex Ginibre enesemble,
see the list of references 
in \cite{Ak1}, \cite{Ak2}, \cite{AkStrahov}. 

{\bf Definition}.
The {\it expectation} of a function $f$ which depends on entries of the $N\times N$ complex 
matrices $Z$ and its Hermitian conjugate $Z^\dag$ is defined by
\be\label{expect}
\E_{N}(f)=\int f(Z) d\mu(Z).
\ee

{\bf Notations}.
 For any given matrix $X$  and a partition 
$\Delta=(\Delta_1,\Delta_2,\dots,\Delta_\ell)$ we introduce the following
notations
\be\label{p(X)'}
\bpow(X)=\left(\tr X , \tr (X^2) , \tr (X^3),\dots \right)
\ee
\be\label{p-lambda-(X)}
{\bf {p}}^*_\Delta (X)=\tr \left(X^{\Delta_1}\right)  \tr \left(X^{\Delta_2}\right) \cdots \tr 
\left(X^{\Delta_\ell}\right)
\ee
Each $\tr X^{m}=:p_{m}^*(X)$ is the so-called {\it power sum} \cite{Mac} (or, the same, {\it Newton sum}) of the 
eigenvalues  of the $N\times N$ matrix $X$. 

In particular, we get 
$$\bpow(\mathbb{I}_N)=\left(N , N , \dots \right)$$ 
and, therefore,
\be\label{p*(I_N)}
\bpow_\Delta(\mathbb{I}_N)=N^{\ell(\Delta)}
\ee
where $\mathbb{I}_N$ is $N\times N$ identity matrix and $\ell(\Delta)$ is the length of $\Delta$
(the {\it length} of a partition is the number of it's non-vanishing parts).

\paragraph{Independent Ginibre ensembles.}
On this subject there is an extensive literature, for instance see \cite{Ak1,Ak2,AkStrahov,S1,S2}.

We will consider integrals over $N\times N$ complex matrices $Z_1,\dots,Z_n$ where the measure is defined as
\be\label{CGEns-measure}
d\Omega(Z_1,\dots,Z_n)= \prod_{\alpha=1}^n d\mu(Z_\alpha)=c_N^n 
\prod_{\alpha=1}^n\prod_{i,j=1}^N d\Re (Z_\alpha)_{ij}d\Im (Z_\alpha)_{ij}\text{e}^{-N|(Z_\alpha)_{ij}|^2}
\ee
where the integration range is $\mathbb{C}^{N^2}\times \cdots \times\mathbb{C}^{N^2}$ and where $c_N^n$
is the normalization 
constant defined via $\int d \Omega(Z_1,\dots,Z_n)=1$.

The set of $n$ $N\times N$ complex matrices
and the measure (\ref{CGEns-measure}) is called {\it $n$ independent complex 
Ginibre ensembles}. 

The expectation of a quantity
$f$ which depends on entries of the matrices $Z_1,\dots,Z_n$ is defined by
$$
\E_{\mathbb{GL}_N^{\otimes n}}(f):=\int f(Z_1,\dots,Z_n) d\Omega(Z_1,\dots,Z_n).
$$
The subscript $ n $ reminds that the expectation is estimated in the product of $ n $ independent
Ginibre ensembles, and the second subscript, $ N $, - that the Gauss measure is not chosen as
$ e^{- \tr ZZ^\dag} $, but in the form $ e^{-N \tr ZZ^\dag} $.

\paragraph{Spectral correlation functions.} We recall that for a given matrix $X$ and a partition 
$\Delta=(\Delta_1,\Delta_2,\dots,\Delta_\ell)$ we introduced (see \ref{p-lambda-(X)}) the following
notation
$$
{\bf {p}}^*_\Delta (X)=\tr \left(X^{\Delta_1}\right)  \tr \left(X^{\Delta_2}\right)
\cdots \tr \left(X^{\Delta_\ell}\right)
$$
where each $\tr \left(X^{\Delta_i}\right)$ is the Newton sum $\sum_{a=1}^N x_a^{\Delta_i}$ of the eigenvalues 
$x_a,\,a=1,\dots,N$ of the matrix $X$.

We are interested in the spectral correlation functions
$\E_{\mathbb{GL}_N^{\otimes n}}(\bpow_{\Delta^1}(X_1)\cdots \bpow_{\Delta^m}(X_m))$ where $X_i,\,i=1,\dots,m$
is a set of matrices and 
$\Delta^i=(\Delta^i_1,\Delta^i_2,\dots),\,i=1,\dots,m$ 
is a set of given partitions.

Let us introduce the notations $\bpow=(p_1,p_2,\dots)$.

 The notations are
as follows: $\bpow_\Delta=p_{\Delta_1}p_{\Delta_2}\cdots$, and 
$z_\Delta = \prod_{i=1}^\infty i^{m_i}m_i!$ where $m_i$ is the number of parts
$i$ which occur in the partition $\Delta$. For instance, for the partition $\Delta=(5,5,2,1,1)$
we get $z_\Delta=5^2 \times 2!\times 2 \times 1!\times 1^2\times 2!  = 200$.

\begin{Remark} \label{Taylor-for-prod-e^V}
Let us note that the generation function of the spectral invariants may be choosen as
\be
\E_{\mathbb{GL}_N}
\left( e^{N\tr \texttt{V}(X_1,\bpow^{(1)})}\cdots e^{ N \tr \texttt{V}(X_\f,\bpow^{(\f)})}  \right)
\ee
Indeed, with the help of (\ref{Taylor-for-e^V})  the Taylor series in parameters ${p}_k^{(i)}$ 
yields the mentioned spectral correlation functions.
\end{Remark}

In what further the matrices $X_i$ will be the dressed words defined in the previous section.

\paragraph{Unitary matrices.}

In what follows we will consider also ensembles of $n$ unitary matrices which are 
the set of $U_i\in\mathbb{U}_N$ and of Haar measures $d_*U_i$. The expectation of a
function of entries of matrices $U_i,\, i=1,\dots,n$ are defined as
\[
 \E_{\mathbb{U}_N^{\otimes n}} (f) = \int f(U_1,\dots,U_n)\prod_{i=1}^n
d_*U_i
\]

\paragraph{Mixed ensembles.}

We also consider mixed ensembles which consist of $n_1$ unitary matrices $U_1,\dots,U_{n_1}$,
with the Haar measures $d_*U_1,\dots,d_*U_{n_1}$ and $n_2$ complex matrices $Z_1,\dots,Z_{n_2}$
with the Gauss measure $d\mu(Z_1),\dots,d\mu(Z_{n_2})$. In the mixed ensemble the expectaion 
of a function of the entries of $n=n_1+n_2$ matrices $U_1,\dots,U_{n_1},Z_1,\dots,Z_{n_2}$ is defined as
\be\label{mixed}
\E_{\mathbb{U}_N^{\otimes n_1}}\E_{\mathbb{GL}_N^{\otimes n_2}} (f) 
 =\int f(U_1,\dots,U_n,Z_1,\dots,Z_{n_2})\prod_{i=1}^{n_1}
d_*U_i\prod_{i=1}^{n_2}d\mu(Z_i)
\ee
where the normalization is chosen to be $\E_{\mathbb{U}_N^{\otimes n_1}}\E_{\mathbb{GL}_N^{\otimes n_2}} (1)=1$,
and
where we assume that the integral does not depend on the integration order. (Say, it is correct 
for the polynomial functions $f$.)

\section{Integrals of tau functions \label{Integrals-of-tau-functions}}

\subsection{Words and dual words}

\paragraph{Decorated embedded graph. Words and dual words.} 
Consider an alphabet consisting of the characters $ A_i $ and $ B_i $, $ i = 1, \dots, n $. Symbols $ A_i, B_i $
we call dual pair for each $i=1,\dots,n$, and we call this alphabet the alphabet of pairs.

Next, we consider the oriented compact 
 surface $ \Sigma $ without boundaries with a given embedded graph with $ \f $ faces, $ n $ edges and $ {\V} $ vertices.
We assume that the complement to the graph on $ \Sigma $ is the union of the disks, therefore
the Euler characteristic of $ \Sigma $ is $ \e = \f-n + {\V} $.

Let us decorate the graph as follows. 
First, we number each edge, and place $A_i$ and $B_i$ from both sides of the edge number $i$ in any fixed way.
Second, we number all faces by $c=1,\dots,\f$. 
Let us go around the boundary of the face number $c$ in the clockwise direction and assign to each
boundary edge, say $e_i$, either the symbol $A_i$ in case the edge is directed positively, or $B_i$ in case
the edge is directed negatively. We get the formal product of symbols written from the left to the right
according to the clockwise round trip. This product
defined up to the cyclic permutations of the characters in it we call the {\it word} $W_c$. Thus, each symbol of 
the collection
$\{A_i,B_i,\, i=1,\dots, n  \}$ is assigned to an edge  and to 
each word is assigned to a face. 

Notice that each symbol of the alphabet of pairs enter once and only once in the set of words.

Such graph we call {\it decorated} and is denoted $\left(\Gamma,W_1,\dots,W_\f\right)$.
The given full set of words $W=(W_1,\dots,W_\f)$ gives rise to the set of {\it dual words}
$W^*=(W_1^*,\dots,W_{\V}^*)$ as follows:

Let us enumerate the vertices $i=1,\dots,\V$.
Let us go in the counterclockwise direction around a given vertex and (from the left to the right) write down symbols
which we meet aprior each outcoming edge. This product defined up to cyclic permutations of the characters in 
the product we call dual word $W_i^*$ assigned to the vertex number $i$. 

We recall that the graph $\Gamma^*$ dual to $\Gamma$ has $\V$ faces, $n$ edges and $\f$ vertices,
where each face of $\Gamma^*$ contains a single vertex of $\Gamma$ and each face of $\Gamma$ contains
a single vertex of $\Gamma^*$. Each edge of $\Gamma$, say $e_i$, crosses a single edge of $\Gamma^*$,
denoted as $e_i^*$. We assign the orientation to each $e_i^*$ in a way that $e_i$ cross it form the left 
to the right. The rule to assign the word to the dual graph is the same, however, now we write the words
from the right to the left. The collection of the words of the dual graph obtained in such a way
coincides with $W_1^*,\dots,W_{\V}^*$ presented above. We have one-to-one correspondence
$$
W_1,\dots,W_\f\quad \leftrightarrow \quad W_1^*,\dots,W_{\V}^*
$$

The decorated graph dual to $\left(\Gamma,W_1,\dots,W_\f\right)$ is denoted 
$\left(\Gamma^*,W_1^*,\dots,W_{\V}^*\right)$.

One can start from the {\it alphabet of pairs} $A_i,B_i,\, i=1,\dots,n$ and the set of products $W_1,\dots,W_\f$
where each letter is used only once. Then one can consider the set of $\f$ polygons: the word 
$W_c,\, c=1,\dots,\f$ gives rise 
the polygon number $c$ whose edges are labeled in clockwise direction by symbols of the word
read from the left. Then, by gluing each pair of edges labeled by a pair $A_i,B_i,\,i=1,\dots,n$
we obtain the decorated embedded graph $\left(\Gamma,W_1,\dots,W_\f\right)$ and also
$\left(\Gamma^*,W_1^*,\dots,W_{\V}^*\right)$.

We call sets of words isomorphic if one can be obtained from another
by transpositions of dual pairs ${a}_i \leftrightarrow {b}_i $, by a re-enumaration of edges
and by re-enumaration of faces.

\paragraph{Chord diagrams and decorated chord networks.} 
In the previous paragraphs, words were introduced with the help of a decorated graph drawn on a Riemann surface
$\Sigma$.

In this paragraph, we ``forget'' about the surface and the embedded graphs and treat a word simply 
as a product of characters 
where each product is defined up to cyclic permutations. As before, we consider 
the alphabet of pairs, which consists of $n$ pairs of dual characters ${a}_i,{b}_i,\,i\in\textbf{I}$, and require that each 
character enter only once in the word set $w=({W}_1,\dots,{W}_\f)$.
We call the set of words {\it connected} if there is no a subset of words which contains only characters
from a subalphabet of pairs. 

A set of words can be drawn on the list of paper in a natural way as a set of $\f$ oriented polygons 
with the total number of edges equal to $2n$ , 
where symbols are assigned to
edges of the polygons: each word is obtained by going clockwise
around the related to the word polygon and multiplying
from left to right of all the characters along the way. 
We draw lines which we call {\it chords} whose endpoints are placed on the edges
with dual symbols.

For further purposes, it is convinient to draw not single chords {\it dual chords}: two directed arrows that together
with arrows $A_i$ and $B_i$ form (topologically) a 4-poligon as follows. Let the arrow $A_i$ start at the point
1 and end at the point 2, and the arrow $B_i$ start at point 3 and end at the point 4, then the chord $A_i^*$
starts at the point 1 and ends at the point 4 and the chord $B_i^*$ starts at the point 3 and ends at the point 2.
Notice that arrows-characters and arrows-chords are directed oppositely on the polygon 1234: both characters are 
directed positevely and both chords are directed negatively.

Let us call this set decorated set of polygons.

We recall that a {\it chord diagram} is an oriented circle $ S^1 $ with a certain
the number of pairs of points connected by lines called chords. A {\it network} of chord diagrams is a set of oriented loops
and a set of lines, also called chords, each chord connecting a pair of points belonging to any circle.
We call a chord internal if its endpoints belong to the same circle, and otherwise we call it external.

Decorated sets of polygons (networks) we call isomorphic if they correspond to the isomorphic sets of words.

Below we consider only connected networks. Thus, we also ask the set of words to be connected
(which means that there is no the subset of words constructed with the  subalphabet of pairs).

Consider a network with $\f$ loops and $n$ chords.
One can naturaly dentify such network with the set of $\f$ polygons with $n$ pairs of edges, where edges 
polygons are segments of loops which contains endpoints of the chords.

 Let us numerate chords and decorate the network assigning 
matrices ${a}_i$ and ${b}_i$ to the segments of loops containing the endpoints of the $i$-th chord. 
Then, up to $\mathbb{Z}_2^n$ action $A_i \leftrightarrow {b}_i,\, i=1,\dots,n$ we have the one-to-one correspondence
between the set of words $w=({W}_1,\dots,{W}_{\f})$ and the decorated network or, the same, to the decorated
set of polygons.

The well-known fact (see, for instance \cite{ZL}) is that any set of oriented polygons with
the total number of (the oriented) edges equal to $2n$ gives rise to an embedded graph drawn on a
Riemann surface. It is obtained by the identification of the pair of oriented edges in such a way
that the origin of one oriented edge coincides with the end point of the other. 
Theses identified oppositely oriented edges of polygons turn to be edges of the embedded graph.

A connected set of words gives rise to the connected decorated graph drawn on the connected Riemann
surface and, therefore to the dual set of words: 
\be\label{geometric-construction-of dual-words}
{W}=({W}_1,\dots,{W}_\f)\to (\Gamma, {W}_1,\dots,{W}_\f)  \leftrightarrow (\Gamma^*, W^*_1,\dots,W^*_{\V}) 
\ee

\paragraph{Operations $\mathfrak{m}_i$ and H(i)  with words and networks.}
Consider the decorated network as the symmetric tensor
 product of $ {W}_1 \otimes \cdots \otimes {W}_\f $.
Select any pair $ A_i, B_i, \, i \in \textbf {I} $. This pair is either 
in different words, say ${W}_a=A_iX_i$ and ${W}_b=B_iY$, or in one word, say $ {W}_c= A_iX_iB_iY $.
Introduce the involutive map $ \mathfrak{m}_i$ which acts on 
the tensor products of words: it acts identically on all words except these (this) that contain(s) 
symbols $A_i$ and $B_i$
as follows:
\be\label{mapL=m1}
 \mathfrak{m}_i:\quad 
A_iX \otimes B_i Y\,\to\, A_iX B_iY
\ee
in the first case, and as
\be\label{mapL=m2}
 \mathfrak{m}_i:\quad    
 A_iXB_iY\,{\to}\, A_iX \otimes B_iY
\ee
in the second case. 
One should pay attention to the 
coordination of the order of factors on the left and right sides of the maps 
(\ref{mapL=m1}) and (\ref{mapL=m2}) and remember that
thanks to the fact that words are defined up
to the cyclic permuations of the characters,
the left hand side of (\ref{mapL=m1}) can be also written as 
$XA_i \otimes YB_i = A_iX \otimes YB_i  = XA_i \otimes B_iY$, and the left hand side of (\ref{mapL=m2})
can be written as $Y A_iXB_i = B_iY A_iX=  XB_iYA_i$.

We see that $\mathfrak{m}_i^{2}$ 
is the identity map. One can check that 
$\mathfrak{m}_i\left(\mathfrak{m}_j(W)\right)=\mathfrak{m}_j\left(\mathfrak{m}_i(W)\right),\,i,j=1,\dots n$.

We recall that having the set of words $W$ and polygons we construct the Riemann
surface with the decorated graph in a unique way, and then we have the geometric construction for
the dual set of words ${W}^*$, see (\ref{geometric-construction-of dual-words}). It was the geometric
construction of dual words.

The algebraic construction is given by

\bp\label{algebraic-construction-of dual-words}
\[
\left(\prod_{i=1}^n \mathfrak{m}_i \right)  {W}_1 \otimes \cdots \otimes {W}_\f =
 {W^*_1} \otimes \cdots \otimes {W^*_{\V}}
\]
\ep

As for the decorated network of chord diagrams, each operation $ \mathfrak{m}_i, \, i \in \textbf{I} $
means the following. We represent the chord and a pair of directed edges, on which it rests, in the form of the 
letter H, where the middle line denotes the chord, and the directionality of the edges is depicted as
$ {1 \atop 2} \downarrow - \uparrow {3 \atop 4} $. Operation
$\mathfrak{m}_i$ means the transposition of the endpoints 2 and 3:
 now new directed edges connect not points 12 and 43, but points 13 and 42 (notice that points 1 and 4 
are origins of the directed edges in both cases). And 
these new edges are connected by the new chord (the middle line of the letter H lying on its side).
One may call it H-rotation, or, H(i)-rotation having in mind that the chord and the edges are numbered by $i$.

The proof of the Proposition \ref{algebraic-construction-of dual-words} is based on the realization of the fact 
that composition of
H(1),\dots,H(n) describes the way from the decorated graph $\Gamma$ to the dual decorated graph $\Gamma^*$.
It is clear because each vertex of the ribbon graph is the face of the graph formed by punctured arrows,
while the faces of the ribbon graph $\Gamma$ are the vertices of the dual graph.

\begin{Example}\label{razvertka-Gamma}
$W= A_1A_2B_1B_2\cdots A_{n-1}A_n B_{n-1}B_n$. Then $W^*=A_2A_1B_2B_1 \cdots A_{n}A_{n-1} B_{n}B_{n-1} $.  
\end{Example}\label{E=2}
\begin{Example}\label{E=2..}
 $W= A_1B_1 A_2B_2\cdots A_n B_n$. Then $W^*=A_1\otimes\cdots \otimes A_n\otimes B_nB_{n-1}\cdots B_1  $
\end{Example}

\begin{Example}
 $W= A_1\cdots A_nB_n\cdots B_1$. Then $W^*=    A_n \otimes B_n A_{n-1} \cdots
 \otimes B_3 A_2 \otimes B_2A_1 \otimes B_1$
\end{Example}

\paragraph{The surface} $\Sigma_{\hc,\mc}$. 
Let us remove $\hc$ pairs of faces of the decorated embedded graph $\Gamma$ and glue each pair by a handle.

Next, we remove
$\mc$ faces of $\Gamma$ and glue M\"{o}bius strips to the boundary of these faces.
 The surface of the Euler
characteristic $\e-2\hc-\mc$ obtained from $\Sigma$ by the manipulation described above we denote $\Sigma_{\hc,\mc}$.
The complement to the graph $\Gamma$ drawn of this surface is a union of $\f-2\hc-\mc$ discs, $\hc$ cylinders and
$\mc$ M\"{o}bius strips.

We will denote the words on the boundary of cylinders as $W_i^+,W_i^-,\,i=1,\dots ,\hc$, where $i$ is the number of 
the handle. The words on the boundary of M{\"o}bius we will denote $W_i',\,i=1,\dots,\mc$.

Thus, in our new notations the set of all words of the decorated $\Gamma$ consists of 
$W_i^+,W_i^-,\,i=1,\dots ,\hc$, $W_i',\,i=1,\dots,\mc$ and the set
$W_1,\dots,W_{\f-2\hc-\mc}$ which coresponds to the faces which we do not remove.

$\quad$

\paragraph{Dressing and dressed words.}  Consider any function $f$ 
of matrices $A_i,B_i,\,i=1,\dots,n$.
One can split the index set $\textbf{I}=\{ i=1,\dots,n\}$ into two groups: 
$\textbf{I}=\textbf{I}_1\cup \textbf{I}_2$. One defines \textit{dressed} function which we denote 
$\Lu^{\textbf{I}_1} \Lz^{\textbf{I}_2}(f),\,c=1,\dots,\f$  by
the following replacment:
\[
A_i,B_i \, \to \, U_iA_i, U^\dag_iB_i\quad {\rm if}\quad i\in\textbf{I}_1 
\]
and
\[
A_i , B_i \, \to \, Z_iA_i,Z_i^\dag B_i \quad {\rm if}\quad i\in\textbf{I}_2
\]
in $f$. In particular, dressed words are denoted $\Lu^{\textbf{I}_1} \Lz^{\textbf{I}_2}\left(W_c\right)$.

\subsection{Integrals of products of Schur functions}

\bp Consider  a set of partitions $\lambda^1=\lambda,\lambda^2,\dots,\lambda^\f $ and the set of 
words $W=(W_1,\dots,W_\f)$. Suppose the set $1,\dots,n$ is splitted into two sets $I_1$ and $I_2$,
$|I_i|=n_i,\,i=1,2$.
We get
\be
\E_{\mathbb{\mathbb{U}}_N^{\otimes n_1}}
\E_{\mathbb{GL}_N^{\otimes n_2}}
\left( \Lu^{I_1} \Lz^{I_2} \left( \prod_{i=1}^{\f} s_{\lambda^i}\left({W}_{i} \right)\right)\right)
 =\delta_\lambda
\frac{\left(|\lambda| ! \right)^{n_2} }
{N^{n|\lambda|}}
\left({\rm dim}_{\mathbb{GL}}\lambda  \right)^{-n_1} 
\left( {\rm dim}_S\lambda  \right)^{-n_2}  \prod_{i=1}^{\V} s_{\lambda}\left(W^*_i\right)
\ee
and
\be
\E_{\mathbb{\mathbb{U}}_N^{\otimes n_1}}
\E_{\mathbb{GL}_N^{\otimes n_2}}
\left( \Lu^{I_1} \Lz^{I_2} \left( \prod_{i=1}^{\V} s_{\lambda^i}\left({W^*}_{i} \right)\right)\right)
 =
\delta_\lambda
\frac{\left(|\lambda| ! \right)^{n_2} }
{N^{n|\lambda|}}
\left({\rm dim}_{\mathbb{GL}}\lambda  \right)^{-n_1} 
\left( {\rm dim}_S\lambda  \right)^{-n_2}  \prod_{i=1}^{\f} s_{\lambda}\left(W_i\right)
\ee
\ep

{\bf Proof.} We notice that according to (\ref{s(ZAZB)}) and (\ref{s(ZA)s(ZB)}) the integration
over $Z_i$ may be described as the action of $m_i$ in (\ref{mapL=m1}) and (\ref{mapL=m2}) respectively,
and the integration over all matrices is described by Proposition \ref{algebraic-construction-of dual-words}.

\subsection{Tau functions and their diagonal parts}
For our purposes of evaluation of integrals of tau functions (\ref{tau-F-KP'}),(\ref{A-prefactor}) 
it is suitable to consider 
even-component KP hierarchy - one can freeze the dependence on, say, $\bpow^{2\textsc{d}}$ and 
get ($2\textsc{d}-1$)-component tau function.
Let us also  
 re-define the higher times with even numbers as follows $\bpow^{2i}\to -\bpow^{2i}$ (Actually, it is natural
redefinition in case we start not from the $2\textsc{d}$-component KP, but from $\textsc{d}$-component 
Toda lattice). We have in mind the usage of (\ref{reflection}). After the replacement and using the Frobenius
coordinated for partitions we get
(\ref{tau-F-KP'}) as
\be\label{tau-F-KP}
\tau^\textbf{A}(\bpow^{1},\dots,\bpow^{2\textsc{d}})=1+
\ee
\[
\sum_{\alpha^1,\dots,\alpha^\textsc{d}\atop \alpha^1,\dots,\alpha^\textsc{d} }
\textbf{A}_{(\alpha^1|\beta^1),\dots,(\alpha^\textsc{d}|\beta^\textsc{d})}\,
 s_{(\alpha^1|\beta^1)}(\bpow^1)s_{(\beta^2|\alpha^2)}(\bpow^2)\cdots
 s_{(\alpha^{2\textsc{d}-1}|\beta^{2\textsc{d}-1})}(\bpow^{2\textsc{d}-1})
 s_{(\beta^{2\textsc{d}}|\alpha^{2\textsc{d}})}(\bpow^{2\textsc{d}})
\]
where $\textbf{A}_{(\alpha^1|\beta^1),\dots,(\alpha^\textsc{d}|\beta^\textsc{d})}$ is given by (\ref{A-prefactor})
\be\label{A-prefactor-1}
\textbf{A}_{\lambda^1,\dots,\lambda^{2\textsc{d}}}
=\det 
\left(\textbf{A}^{i,j}_{\alpha^{(i)}_{l^{(i)}},\beta^{(j)}_{m^{(j)}} }
\right)_{i,j=
1,\dots,\textsc{d},\,l^{(i)}=1,\dots,\kappa^{(i)},m^{(j)}=1,\dots,\kappa^{(j)} }
\ee
Now let us consider the projection of the series over set of $2\textsc{d}$ partitions on its diagonal part,
i.e. on the subsum where all partition are equal:
\be\label{projection}
\Pi 
\tau^\textbf{A}(\bpow^{1},\dots,\bpow^{2\textsc{d}})=
\sum_\lambda A(\lambda)\prod_{i=1}^{\textsc{d}} s_\lambda(\bpow^{2i-1})s_\lambda(-\bpow^{2i})
\ee
where 
\be\label{A(lambda)}
\textbf{A}(\lambda)=\textbf{A}_{\lambda,\dots,\lambda},
\ee
$\lambda=(\alpha|\beta)$ and
$\textbf{A}(\lambda)$ is symmetric separetely in the set of $\alpha_i$ and the set of $\beta_i$.
Indeed, after the projection we get $\alpha^{2i-1}=\beta^{2i}=\alpha$ and $\alpha^{2i}=\beta^{2i-1}=\beta$
in formula (\ref{A-prefactor-1}). 
It means that if we permute $\alpha_i$ and $\alpha_j$ we sumalteneousely permtue the same number of pairs
of rows and columns and the determinant (\ref{A-prefactor-1}) will be the same. The same is true for the 
parmutations of parts of $\beta$.

Let us consider examples.

 Consider tau function

\be\label{tau-F-KP-example}
\tau^\textbf{A}(\bpow^{1},\dots,\bpow^{2\textsc{d}})=1+
\ee
\[
\sum_{\alpha^1,\dots,\alpha^\textsc{d}\atop \alpha^1,\dots,\alpha^\textsc{d} }
\prod_{i=1}^{2\textsc{d}}
{A}_{(\alpha^i|\beta^i)}\,\prod_{i=1}^{\textsc{d}}
 s_{(\alpha^{2i-1}|\beta^{2i-1})}(\bpow^{2i-1})s_{(\beta^{2i}|\alpha^{2i})}(\bpow^{2i})
\]
where $A$ is some matrix. We get 
$\textbf{A}(\lambda)=
\det\left( A_{\alpha_i,\alpha_j}  \right)_{i,j}\det\left(A_{\beta_i,\beta_j}  \right)$

For $A_{i,j}=\delta_{i,j}e^{\sum_{m>0} t_m( i+\frac12)^m}$, 
\be\label{comp-cycle-prefactor}
\textbf{A}(\lambda)=
e^{\sum_i\sum_{m>0} t_m ( (\alpha_i +\frac 12 )^m -(-\beta_i-\frac 12 )^m)}
\ee

\subsection{Tau functions as integrands}

Let use the following convention similar to (\ref{convention}): 
If a given set of higher times, say, $\bpow^i$ is specified as
Newton sums of the eigenvalues if certain matrix, say $X_i$, then instead of the variable $\bpow^i=\bpow^i(X_i)$
we write capital $X_i$. By ${\rm dim}_{\mathbb{GL}}\lambda=s_\lambda(\mathbb{I}_N)$ and 
${\rm dim}_{S}\lambda=\chi_\lambda(1^d)$ 
we denote dimensions of irreducible representations $\lambda$ of linear and symmetric groups resectively.

\paragraph{Orientable case.}

\bp\label{int-tau-A} Consider tau function (\ref{A-prefactor}) where higher times are given by 
\be\label{p=p(W)}
\bpow^i = \bpow^i(W_i) = \left(p^i_1(W_i), p^i_2(W_i), p^i_3(W_i),\dots \right),\quad 
p^{(i)}_m=\tr\left((W_i)^m\right)\quad
i=1,\dots,2\textsc{d}
\ee
Then
\be
\E_{\mathbb{\mathbb{U}}_N^{\otimes n_1}}
\E_{\mathbb{GL}_N^{\otimes n_2}}
\left( 
\Lu^{I_1} \Lz^{I_2} \left(
\tau^{\textbf{A}}(W_1,\dots,W_{2\textsc{d}})\right)
\right)
\ee
\be\label{sum-A}
 =1+\sum_{d>0}\sum_{\lambda\atop |\lambda|=d}
\frac{\left( d ! \right)^{n_2} }
{N^{nd}} \textbf{A}(\lambda)
\left({\rm dim}_{\mathbb{GL}}\lambda  \right)^{-n_1} 
\left( {\rm dim}_S\lambda  \right)^{-n_2}  \prod_{i=1}^{\V} s_{\lambda}\left(W^*_i\right)
\ee
where $\textbf{A}(\lambda)$ is given by (\ref{A(lambda)}),  it is a symmetric function of 
$\alpha_1,\dots,\alpha_{\kappa}$ and of $\beta_1,\dots,\beta_{\kappa}$, where
$\lambda =(\alpha|\beta)$
\ep 

 Consider the integrals of the tau function given by (\ref{comp-cycle-prefactor})
\[
\E_{\mathbb{\mathbb{U}}_N^{\otimes n_1}}
\E_{\mathbb{GL}_N^{\otimes n_2}}
\left( 
\Lu^{I_1} \Lz^{I_2} \left(
\tau^{\textbf{A}}(W_1,-\bpow^2,W_2,-\bpow^4,\dots,W_{2\textsc{d}-1},-\bpow^{2\textsc{d}})\right)
\right)=
\]
 \be\label{witten}
\sum_\lambda 
\frac{\left(|\lambda| ! \right)^{n_2} }
{N^{n|\lambda|}}
e^{\sum_i\sum_{m>0} t_m ( (\alpha_i+\frac 12)^m -(-\beta_i-\frac 12)^m)} 
\left({\rm dim}_{\mathbb{GL}}\lambda  \right)^{-n_1} 
\left( {\rm dim}_S\lambda  \right)^{-n_2} 
\prod_{i=1}^{\textsc{d}} s_\lambda(\bpow^{2i})\prod_{i=1}^\V s_\lambda(W^*_i)
\ee
where the sum ranges over all partitions $\lambda=(\alpha|\beta)$ whose length does no exceed $N$.
In case $t_m=\delta_{2,m}$ and $n_2=0$ this expression coinsides with formula (2.79) in \cite{Witten}
for the correlation function of $\f+\V$ Wilson loops on the orientable surface of genus $ \f+\V -n$.

\bp
Let $\e=D+\V-n=2$ and each $\bpow^{j}$ except two ones from the set 
$\bpow^{2i},\bpow^{2i-1}(W_i)$ can be presented in form (\ref{special-p}). Then
the sum (\ref{witten}) is two-component KP tau function of hypergeometric type \cite{OS-2000}.
\ep

\paragraph{Non-orientable case.}

Consider the simplest nontrivial tau functon of the one component BKP which is
\be
\tau^B(\bpow) =\sum_\lambda s_\lambda(\bpow)
\ee
Then, we get
\bp\label{int-tau-Klein}
\be
\E_{\mathbb{\mathbb{U}}_N^{\otimes n_1}}
\E_{\mathbb{GL}_N^{\otimes n_2}}
\left( 
\Lu^{I_1} \Lz^{I_2} \left(
\tau^A(W_1,\dots,W_{2\textsc{d}})\right)\tau^B(W_{2\textsc{d}+1})
\right)
\ee
\be\label{sum-B}
 =\sum_{\lambda\atop \ell(\lambda)\le N}
\frac{\left(|\lambda| ! \right)^{n_2} }
{N^{n|\lambda|}} A(\lambda)
\left({\rm dim}_{\mathbb{GL}}\lambda  \right)^{-n_1} 
\left( {\rm dim}_S\lambda  \right)^{-n_2} s_\lambda(\bpow^1) \prod_{i=1}^{\V} s_{\lambda}\left(W^*_i\right)
\ee
where $\textbf{A}(\lambda)$ is the same as in Proposition \ref{int-tau-A}.
\ep

\subsection{Discrete $\beta$-ensembles}

Sums in the right hand sides of (\ref{sum-A}) and  (\ref{sum-B}) 
may be treated as discrete ensembles which
generalize known ensembles which can be related to $\e$ series in the Schur functions 
\cite{KMMM} and \cite{OShiota-2004}. 

$\,$

\paragraph{$\beta$-ensemble.\label{Section-beta-ensembles}}
The matrix models labelled with networks may written as discrete $\beta$-ensembles if we fix parameters
$\bpow^{(c)}$ with the help (\ref{special-p}) that means that we study expectation value of products
of powers of determinants (and one of this power should be a natural number. This topic will be developed 
in a more detailed version, now, let me explain the idea. One need to use relations
\be
s_\lambda(N\bpow(\texttt{d},a))= a^{|\lambda|}(-N\texttt{d})_\lambda
s_\lambda(\bpow_\infty) ,
\quad s_\lambda(N\bpow_\infty)= N^{|\lambda|}
\frac{\prod_{i<j}^N(h_i-h_j)}{\prod_{i=1}^N h_i!}
\ee
where $h_i=\lambda_i-i+N$ are shifted parts of $\lambda$ and the notation $(-\texttt{d})_\lambda $
was defined in (\ref{Poch-YD}). Let $N\texttt{d}_1=NL>0$ is integer.
Notice that $(-NL)_\lambda $ vanishes
for $\lambda_1>NL$. For $N\texttt{d}_i$ that are not natural numbers, we use
\[
 (-N\texttt{d}_i)_\lambda =\prod_{j=1}^{N-1} (-N\texttt{d}_i-j)^{N-j+1} 
 \prod_{j=1}^{N} \frac{\g(h_j+1-N-N\texttt{d}_i)}{\g(-N\texttt{d}_i)}
\]
Then, choosing any $e$ within $0 \le e \le {\f}-1$,  we get
\[
 \E_{\mathbb{GL}_N^{\otimes n}}\left(  
 \det \left(1-a_1\Lz(W_1)\right)^{NL} 
 \prod_{i=2}^{{\f}-1-e}
 \left(1-a_i\Lz(W_i)\right)^{N\texttt{d}_i}
 \prod_{i={\f}-e+1}^{\f}{\tau_1^{\rm 2KP}}(\Lz(W_i))
 \right)=
\]
\be\label{discrete-beta-ensemble}
=\,\frac{c_N}{N!}\,\sum_{h_1,\dots,h_N\ge 0}' \,\prod_{a<b}^N\, |h_a-h_b|^{{\f}-n+{\V}-e}\,
\prod_{j=1}^N \,\frac{a_1^{h_j}\prod_{i=1}^{{\f}-e-1} a_i^{h_j}
\g(h_j+1-N-N\texttt{d}_i)}{\left(\g(h_j+1)\right)^{{\f}-n+{\V}}\g(NL+N-h_j)}
\ee
where $\Sigma'$ means that all $h_i,\,i=1,\dots,N$ are different (therefore the Vandermond product
does not vanish), and
where $c_N=c_N(\{a_i,\texttt{d}_i\})=\prod_{i=1}^{{\f}-e}
a_i^{...}
\prod_{j=1}^N 
\frac{(-N\texttt{d}_i-j)^{N-j+1}}
{\g(-N\texttt{d}_i)}....$.

We intentionaly separate the case $N\texttt{d}_1=NL$  to avoid possible
divergence in the summation, with $L$ be a natural number the right hand side (\ref{discrete-beta-ensemble})
it is a finite sum with the summation range $ 0 \le h_i \le NL+N,\,i=1,\dots,N$.

One could write down the equation for the equilibrium Young diagram related to the discrete 2D Coulomb gas 
on the semiline (in case $\beta=1,2$, see \cite{Forrester} for many details) or, 2D 'gravitational' gas on the 
semiline in case $\beta <0 $.

$\,$

\paragraph{Coupled, or, Kontsevich-type ensembles.}
It may be available to fix $\bpow^{(2)}$ in different way as $\bpow^{(2)}=\bpow^{(2)}(Y)$ where
$Y_{ij}=\delta_{i,j}\exp y_i,\,i=1,\dots, N$ (see (\ref{p(X)}) for the notation). 
The matrix $Y$ plays the role of an additional source matrix similar to the role of
external matrix in the coupled matrix model.
(One can still take any of $N\texttt{d}_i$ to be an natural number in case the sum is divergent).
Instead of (\ref{discrete-beta-ensemble}) we get
\[
 \E_{\mathbb{GL}_N^{\otimes n}}\left(  
 \det \left(1-Y \otimes \Lz(W_1)\right)^{-N} 
 \prod_{i=2}^{{\f}-1-e}
 \left(1-a_i\Lz(W_i)\right)^{N\texttt{d}_i}
 \prod_{i={\f}-e+1}^{\f}{\tau_1^{\rm 2KP}}(\Lz(W_i))
 \right)=
\]
\be\label{discrete-kontsevich-ensemble}
\frac{\tilde{c}_N}{N!}\,\sum_{h_1,\dots,h_N\ge 0}' \,\prod_{a<b}^N\, (h_a-h_b)^{{\f}-n+{\V}-e-1}\,
\prod_{j=1}^N \,\frac{a_1^{h_j}\prod_{i=1}^{{\f}-e-1} a_i^{h_j}
\g(h_j+1-N-N\texttt{d}_i)}{\left(\g(h_j+1)\right)^{{\f}-n+{\V}}
\exp (-N y_j h_j)}
\ee
where $\tilde{c}_N=c_N \prod_{i<j} (e^{Ny_i}-e^{Ny_j})$, compare to the similar replacement in \cite{KMMM}
and \cite{OShiota-2004}.

\section*{Acknowledgements}

A.O. is grateful to J. van de Leur for numerous discussions concerning BKP hierarchies, to
E.Strahov who attracted his attention to the products of random matrices, we thanks
 S.Lando and M.Kazarian for useful comments and  A.Gerasimov who paid 
our attention to the paper \cite{Witten}. A.O. is thankful to A.Odzijewich for his kind hospitality in Bialowezie
where this work was done.
The work of A.O. was done
in the framework of the state assignment of P.P. ShirshovInstitute of Oceanology (theme 0149-2019-0002) and Program 
"Nonlinear dynamics" of Russian Academy of Sciences and was partially supported by RFBR grant 18-01-00273a and
also by the  
Russian Academic Excellence Project `5-100'.

\appendix

   \section{Appendices}

\subsection{Integration of the Schur functions}

\paragraph{The expectationl of $s_\lambda(ZAZ^\dag B)$ and  of $s_\lambda(ZA)s_\nu(Z^\dag B)$ }

\bl 
\label{Prop-s(ZAZB)} 
For any $N\times N$ complex matrices $A$ and $B$ we have
\be\label{s(ZAZB)} 
\E_{\mathbb{GL}_N}\left( s_\lambda(ZAZ^\dag B)\right) = N^{-d}\frac{s_\lambda(A) s_\lambda(B)}{s_\lambda(\bpow_\infty)}
\ee
where $d=|\lambda|$.
\el

In case $A$ and $B$ are Hermitian matrices,
the first relation (\ref{s(ZAZB)}) is well-known and written down in textbooks,  
see for instance Example 5 in Section VII, 5 of \cite{Mac}. (The only difference with
the well-known formula is the factor $N^{-d}$
 which results from the fact that we replace the Gauss weight $e^{-\tr ZZ^\dag}$ (see \cite{Mac})
in the definition of the measure $d\mu$ by $e^{-N\tr ZZ^\dag}$, see (\ref{CGE-measure}). 
Then, the factor $N^{-d}$ is obtained by the rescaling of the Schur function $s_\lambda$ which is the homogenious 
polynomial of the weight $d=|\lambda|$). However, thanks to the fact that we can present Schur functions in form 
(\ref{Schur-char-map}) where each $\bpow_\Delta = \bpow_\Delta(X)$ is a polynomial in the entries of matrix $X$.
Then, the both sides of (\ref{s(ZAZB)}) are analitic functions in the entries of the matrices and, therefore, 
(\ref{s(ZAZB)}) is true for $A,B \in GL_N(\mathbb{C})$.

\bl \label{Prop-s(ZA)s(ZB)} 
For any $N\times N$ complex matrices $A$ and $B$ the following equality is correct:
\be\label{s(ZA)s(ZB)} 
\E_{\mathbb{GL}_N}\left( s_\lambda(ZA) s_\nu(Z^\dag B)\right) = 
N^{-d}\delta_{\lambda\nu}\frac{s_\lambda(AB)}{s_\lambda(\bpow_\infty)}
\ee
where $d=|\lambda|$.
\el

\paragraph{The expectation of $s_\lambda(UAU^\dag B)$ and of  $s_\lambda(UA)s_\nu(U^\dag B)$.}

\bp 
\label{Prop-s(UAUB)} 
For any $N\times N$ complex matrices $A$ and $B$ the following two equalities are correct:
\be\label{s(UAUB)} 
\E_{\mathbb{U}_N}\left( s_\lambda(UAU^\dag B)\right) = 
\frac{s_\lambda(A) s_\lambda(B)}{s_\lambda(\mathbb{I}_N )}
\ee
where 
\be\label{Hpp-U}
\E_{\mathbb{U}_N}\left(\,
\bpow^*_\Delta(UAU^\dag B)\,\right)\,
=
\, z_\Delta\, N^{-\ell(\Delta)}\,\sum_{\Delta^a,\Delta^b}\,
 {^U}H_{\mathbb{CP}^1}(\Delta,\Delta^a,\Delta^b)\,
 \bpow^*_{\Delta^a}(A)\,
 \bpow^*_{\Delta^b}(B)
\ee
where the summation in the right hand side ranges over all partitions $\Delta^a, \Delta^b$ of the weight $d=|\Delta|$
and where
\be\label{Hurwtiz-H(*,*,*)-U}
{^U}H_{\mathbb{CP}^1}(\Delta,\Delta^a,\Delta^b) :=
\sum_{\lambda\atop |\lambda|=d}\left(\frac{ {\rm \dim}_{\mathbb{U}} \lambda}{d!} \right)^2
\frac{\varphi_\lambda(\Delta)\varphi_\lambda(\Delta^a)\varphi_\lambda(\Delta^b)}{\left((N)_\lambda\right)^3}
\ee 

\ep

\bl \label{Prop-s(UA)s(UB)} 
For any $N\times N$ complex matrices $A$ and $B$ the following two equalities are correct and equivalent:
\be\label{s(UA)s(UB)} 
\E_{\mathbb{U}_N}\left( s_\lambda(UA) s_\nu(U^\dag B)\right) = 
\delta_{\lambda\nu}\frac{s_\lambda(AB)}{s_\lambda(\mathbb{I}_N)}
\ee
where 
\be\label{Hp-U}
\E_{\mathbb{U}_N}\left(\,
\bpow_{\Delta^a}(UA)\, \bpow_{\Delta^b}(U^\dag B)\,\right)\,=\, 
z_{\Delta^a} z_{\Delta^b}\, N^{-\ell(\Delta^a)-\ell(\Delta^b)}\,
\sum_{\Delta}\,
H_{\mathbb{CP}^1}(\Delta^a,\Delta^b,\Delta)\,\bpow_{\Delta}(AB)
\ee
where the summation in the right hand side ranges over all partitions $\Delta$ of the weight 
$d=|\Delta^a|=|\Delta^b|$
and where
$
H_{\mathbb{CP}^1}(\Delta,\Delta^a,\Delta^b)$  
is the same three-point Hurwitz number with the base $\mathbb{CP}^1$.
\el

\subsection{The sketch of proofs \label{the-sketch-of-proofs}}

We use
\begin{equation}\label{sAZBZ^+'}
\int_{\mathbb{C}^{N^2}} s_\lambda(AZBZ^+)\text{e}^{-N\operatorname{tr}
	ZZ^+}\prod_{i,j=1}^N d^2Z_{ij}=
\frac{s_\lambda(A)s_\lambda(B)}{s_\lambda(N\bpow_{\infty})}
\end{equation} 
and
\begin{equation}\label{sAZZ^+B'}
\int_{\mathbb{C}^{N^2}} s_\Delta(AZ)s_\lambda(Z^+B) \text{e}^{-N\operatorname{tr}
	ZZ^+}\prod_{i,j=1}^N d^2Z_{ij}= \frac{s_\lambda(AB)}{s_\lambda(N\bpow_{\infty})}\delta_{\Delta,\lambda}\,.
\end{equation}

These relations are used for step-by-step integration (Gaussian in the case of complex matrices).

As we can see, these relations perform the procedure of cutting and joining loops in a network of chord diagrams, 
and also create edges of embedded graph (each edge is a coupled pair of conjugate random matrices). 
Namely, the equation (\ref{sAZBZ^+'}) performs the splitting of the loop $AZBZ^\dag$ into two loops,
$A$ and $B$,
for complex Ginibre ensembles (the resulting equation performs the union of two loops $A$ and $B$ for complex 
Ginibre ensembles. Every time we apply some of the relations (\ref{sAZBZ^+'})-(\ref{sAZZ^+B'}), 
we get the factor (the "propagator" of the edge of the embedded graph), which is 
$ \frac{1}{s_\lambda(N\bpow_\infty)} $
in the case of complex Ginibre ensemble.

\subsection{Hirota equation for the TL and for the 2-component KP tau functions. \label{TL-2-KP}} 
 The TL tau function was introduced in \cite{JM} 
 and may be defined by 
  \be
  \tau^{\rm TL}_n(t,{\bar t}) = \l n|e^{\sum_{i>0} t_i\alpha_i} g^{\rm TL} e^{-\sum_{i>0} {\bar t}_i\alpha_{-i}} |n\r
  \ee

This tau function solves Hirota equation, \cite{JM},\cite{UT}
 \bea\label{Hirota-TL-}
 \oint\frac{dz}{2\pi i}z^{n'-n}e^{\texttt{V}(t'-t,z)}\tau^{\rm TL}_{n'}\left(t'-[z^{-1}],{\bar t}'  \right)
 \tau^{\rm TL}_{n}\left(t+[z^{-1}],{\bar t}  \right)=   \nonumber \\  \oint\frac{dz}{2\pi i} 
 z^{n'-n}e^{V\left({\bar t}'-{\bar t},z^{-1}\right)}\tau^{\rm TL}_{n'+1}\left(t',{\bar t}'-[z]  \right)
 \tau^{\rm TL}_{n-1}\left(t,{\bar t} +[z] \right) 
 \eea
(see \cite{JM}, \cite{UT}) which includes
\be\label{the-first-TL-Hirota}
\frac{\partial^2 \tau^{\rm TL}_n}{\partial t_1\partial {\bar t}_1} \tau^{\rm TL}_n -
\frac{\partial \tau^{\rm TL}_n}{\partial t_1}\frac{\partial\tau^{\rm TL}_n}{ \partial {\bar t}_1} =
-\tau^{\rm TL}_{n+1} \tau^{\rm TL}_{n-1}
\ee

  The two-component KP  tau function
 \be
 \tau^{\rm 2KP}_n(t,{\bar t}) =
 \l n,-n| e^{\sum_{i>0} \left(t_i'\alpha_i^{(1)} +{\bar t}_i'\alpha_i^{(2)}\right)} g^{\rm 2KP} |0\r
 \ee
 solves Hirota equation 
  \bea\label{Hirota-2KP-derivation}
 \oint\frac{dz}{2\pi i}(-)^{-n'-n}z^{n'-n}e^{\texttt{V}(t'-t,z)}\tau^{\rm 2KP}_{n'}\left(t'-[z^{-1}],{\bar t}'  \right)
 \tau^{\rm 2KP}_{n}\left(t+[z^{-1}],{\bar t}  \right) =   \nonumber \\  \oint\frac{dz}{2\pi i} 
 z^{n-n'-2}e^{V\left({\bar t}'-{\bar t},z\right)}\tau^{\rm 2KP}_{n'+1}\left(t',{\bar t}'-[z^{-1}]  \right)
 \tau^{\rm 2KP}_{n-1}\left(t,{\bar t} +[z^{-1}] \right)
 \eea
 which up to the sign factor $(-)^{n+n'}$ in the first integral is (\ref{Hirota-TL-}) if we change $z\to z^{-1}$ 
 in the second integral in (\ref{Hirota-2KP-derivation}).

 \subsection{Pfaffians \label{Pfaffians}}

 If $A$ an anti-symmetric matrix of an odd order its determinant
vanishes. For even order, say $k$, the following multilinear form
in $A_{ij},i<j\le k$
 \be\label{Pf''}
\Pf [A] :=\sum_\sigma
{\sgn(\sigma)}\,A_{\sigma(1),\sigma(2)}A_{\sigma(3),\sigma(4)}\cdots
A_{\sigma(k-1),\sigma(k)}
 \ee
where sum runs over all permutation restricted by
 \be
\sigma:\,\sigma(2i-1)<\sigma(2i),\quad\sigma(1)<\sigma(3)<\cdots<\sigma(k-1),
 \ee
 coincides with the square root of $\det A$ and is called the
 {\it Pfaffian} of $A$, see, for instance \cite{Mehta}.

\section{Hypergeometric BKP Tau Function. Fermionic Formul{\ae}\label{fermionic-appendix}}
Details may be found in \cite{OS-2000, OST-I}.
Let
 $\{\psi_i$, $\psi_i^\dag$, $i \in \mathbb{Z}\}$ be Fermi creation and
annihilation operators that  satisfy the usual anticommutation relations and vacuum annihilation conditions
\[
[\psi_i, \, \psi_j]_+ = \delta_{i,j}, \quad \psi_i | n\r =
\psi_{-i-1} | n\r =0\,,\quad   i< n\,.
 \]
In contrast to the DKP hierarchy introduced in \cite{JM}, for the BKP hierarchy introduced in \cite{KvdLbispec},
we need an additional Fermi mode  $\phi$ which anticommutes with all the other
Fermi operators except itself, for which $\phi^2=1/2$, and
 $\phi|0\r=|0\r/\sqrt{2}$ \cite{KvdLbispec}. Then the hypergeometric BKP tau function introduced in
 \cite{OST-I} may be written as
 \begin{eqnarray}
 g(n)\tau^{\rm BKP}_r(N,n,\bpow) &=&
 \Big\l n\big| e^{\sum_{m>0} \frac 1m J_m p_m}
 e^{\sum_{i < 0} U_i \psi_i^\dag \psi_i -\sum_{i \ge 0} U_i \psi_i\psi_i^\dag }
 e^{\sum_{i>j} \psi_i\psi_j\,-\sqrt{2}\,\phi \sum_{i\in\mathbb{Z}} \psi_i}\big|n-N\Big\r\nonumber \\\noalign{\smallskip}
  &=& \sum_{\lambda\atop \ell(\lambda)\le N} \,e^{-U_\lambda(n)} s_\lambda(\bpow)
 = g(n)
\sum_{\lambda\atop \ell(\lambda)\le N}  r_\lambda(n)s_\lambda(\bpow)\,,\label{hyper-via-U,r}
\end{eqnarray}
where  $J_m=\sum_{i\in\mathbb{Z}}\,\psi_i\psi^\dag_{i+m}$, $m>0$,
$U_\lambda(n)=\sum_{i} U_{h_i+n}$,
$r(i)=e^{U_{i-1}-U_{i}}$,
and
\begin{equation}\label{g(n)}
g(n):=\Big\l n\big|  e^{\sum_{i < 0} U_i \psi_i^\dag \psi_i -\sum_{i \ge 0} U_i \psi_i\psi_i^\dag }
\big|n\Big\r =\left\{\begin{array}{l l}
 e^{-U_0+\cdots -U_{n-1}}& {\rm if}\,\, n>0\,,\\[3pt]
 1& {\rm if} \,\,n=0\,,\\[3pt]
 e^{U_{-1}+\cdots U_{n}}& {\rm if}\,\, n<0\,.
 \end{array}\right.
\end{equation}
In (\ref{hyper-via-U,r}) the summation runs over all partitions whose lengths do not exceed $N$.
\br\label{DKPvsBKP}
Note that, without the additional Fermi mode $\phi$, the summation range in (\ref{hyper-via-U,r}) does
include partitions with odd partition lengths. One can avoid this restriction by introducing a pair of DKP tau
functions, which seems unnatural.
\er
\noindent Apart from (\ref{hyper-via-U,r}), the same series without the restriction $\ell(\lambda)\le N$ gives the BKP tau function. However, it is related to the single value $n=0$. The $n$-dependence destroys the
simple form of this tau function \cite{OST-I}.


\begin{thebibliography}{99}






\bibitem{AvM-Pfaff} M. Adler and P. van Moerbeke, Symmetric random matrices and the
Pfaff lattice, arXiv:solv-int/9903009v1

\bibitem{AMS} M. Adler, P. van Moerbeke and T. Shiota, Pfaff
$\tau$-functions, Mathematische Annalen, 322 (2002) 423--476; arXiv:nlin/9909010


\bibitem{Szabo} M. Aganagic, H. Ooguri, N. Saulina, and C. Vafa, ``Black holes, $q$-deformed 2D Yang--Mills and nonperturbative
topological strings'', Nucl. Phys. B 715(2005) pp. 304--348;
R. Szabo and M.Tierz, ``Chern--Simons matrix models, two-dimensional Yang--Mills theory, and the Sutherland model'',
arxiv preprint hep-th/1003.1228

\bibitem{Ak1} G. Akemann, J. R. Ipsen, M. Kieburg,
\textit{Products of Rectangular Random Matrices: Singular Values and Progressive Scattering},
	arXiv:1307.7560
	
	\bibitem{Ak2} G. Akemann, T. Checinski, M. Kieburg, \textit{Spectral correlation functions of the sum of two independent complex Wishart matrices
	with unequal covariances}, arXiv:1502.01667
	
	\bibitem{AkStrahov} G. Akemann, E. Strahov, \textit{Hard edge limit of the product of two strongly coupled random matrices},
	arXiv:1511.09410
	
	\bibitem{Alexandrov}  Alexandrov, A.: Matrix models for random partitions. Nucl. Phys. B {\bf 851}, 620-650 (2011)
	
	\bibitem{AMMN-2011} A. Alexandrov, A. Mironov, A. Morozov and S. Natanzon,
	\textit{Integrability of Hurwitz Partition Functions.
	I. Summary}, J.Phys.A: Math.Theor.45(2012) 045209,  arXiv: 1103.4100
	
	\bibitem{AMMN-2014} A. Alexandrov, A. Mironov, A. Morozov and S. Natanzon,
	\textit{On KP-integrable Hurwitz functions},JHEP 11(2014) 080,  arXiv: 1405.1395

	
	
	\bibitem{AlexandrovZabrodin-Okounkov} A. Alexandrov and A. V. Zabrodin \textit{Free fermions and tau-functions},
	J.Geom.Phys. 67 (2013) pp. 37-80 ;   arXiv:1212.6049


\bibitem {AN1} Alexeevski A., Natanzon S., Algebra of Hurwitz numbers for seamed surfaces,
Russian Math. Surveys, 61 (4) (2006), 767-769
\bibitem {AN2} Alexeevski A., Natanzon S., Algebra of Bipartite graphs and Hurwitz numbers of
seamed surfaces. Math.Russian Izvestiya  72 (2008) V.4, 3-24.

	
	\bibitem{AN2008} A. V. Alekseevskii and S. M. Natanzon, \textit{The algebra of bipartite graphs and Hurwitz numbers of seamed surfaces},
	Izvestiya Mathematics 72:4 (2008) pp. 627-646

\bibitem{AN3} Alexeevski A., Natanzon S., Hurwitz numbers for regular coverings of surfaces by
seamed surfaces and Cardy-Frobenius algebras of finite groups, Amer. Math. Sos. Transl. (2) Vol 224, 2008, 1-25 (arXiv: math/07093601)

	
	
	\bibitem{AN} A. A. Alexeevski and S. M. Natanzon, \textit{Noncommutative two-dimansional field theories and Hurwitz numbers for real
	algebraic curves}, Selecta Math. N.S. v.12 (2006) ,n.3, pp. 307-377, arXiv:math/0202164

	
	\bibitem{Alfano} G.Alfano,
	``Products of Ginibre and deterministic matrices in the analysis of correlated multiantenna channels''
	https://www2.physik.uni-bielefeld.de
	
\bibitem{Carrell} S.R.Carrell, ``The Non-Orientable Map Asymptotics Constant $p_g$'', arXiv:1406.1760

	
	\bibitem{ChekhovAmbjorn} J. Ambjorn and L. Chekhov \textit{The matrix model
	for hypergeometric Hurwitz number},
	Theoret. and Math. Phys., 1
	81:3 (2014), 1486-1498; arXiv:1409.3553
	
		
	\bibitem{Chekhov-2014}
	J. Ambjorn and L. O. Chekhov, \textit{The matrix model
	for dessins d'enfants},
	Ann. Inst. Henri Poincare D, 1:3 (2014), 337-361;
	arXiv:1404.4240

	\bibitem{BrezinKazakov} E. Brezin and V. Kazakov, \textit{Exactly solvable field theories of closed strings},
	Phys Lett {\bf B236} pp 144-150 (1990);

\bibitem{CZ} L.-L. Chau and O. Zaboronsky,  On the Structure of
Correlation Functions in the normal Matrix Models, {\it Commun.
Math. Phys.} {\bf 196} (1998) 203--247

\bibitem{CGN}  A.F. Costa, S.M. Gusein-Zade S.M. Natanzon  \textit{Klein foams}, Indiana Univ.Math.J. 60 (2011) no 3.
	
\bibitem{DJKM} Date, E., Jimbo, M., Kashiwara, M. and Miwa, T.,
Transformation groups for soliton equations. In:
Jimbo, M. and Miwa, T. (eds) {\it Nonlinear integrable systems---%
classical theory and quantum theory\/} pp. 39--120, World
Scientific, 1983

\bibitem{DJKM-BKP} E. Date, M. Jimbo, M. Kashiwara and T. Miwa,
``Transformation groups for soliton equations, IV A new hierarchy
of soliton equations of KP-type'', Physica 4D (1982) 343-365


\bibitem{Dijkgraaf} R. Dijkgraaf, \textit{Mirror symmetry and elliptic curves, The Moduli Space
of Curves}, R. Dijkgraaf, C. Faber, G. van der Geer (editors), Progress
in Mathematics, 129, Birkhauser, 1995.


\bibitem{D2} Dijkgraaf  R., Geometrical Approach to Two-Dimensional Conformal Field Theory, Ph.D.Thesis (Utrecht, 1989)



	\bibitem{ELSV} T. Ekedahl, S. K. Lando, V. Shapiro and A. Vainshtein,
	\textit{On Hurwitz numbers and Hodge integrals},
	C.R. Acad. Sci. Paris Ser. I. Math. Vol. 146, N2, pp. 1175-1180 (1999)
	
	

	\bibitem{Forrester} P. Forrester, \textit{Log-gases and random matrices}, 2010, Princeton University Press
	
	\bibitem{ForresterWarnaar} P. Forrester and S. Warnaar, \textit{The importance of the Selberg integral}, 2008
	Bulletin of the American Mathematical Society 45 (4), 489-534
	
	

\bibitem{GN} S. M. Gusein-Zade, S. M. Natanzon,
 Klein foams as families of real forms of Riemann surfaces,
 Adv. Theor.Math. Phys. 21(2017), no. 1, 231-241

	\bibitem{GMMMO} A.Gerasimov, A. Marshakov, A. Mironov, A. Morozov and A. Orlov,
	``Matrix models of two-dimensional gravity and Toda theory'', Nuclear Physics B 357 (2-3), 565-618
	 (1992)
	
	
	\bibitem{Goulden-Jackson-2008} I. P. Goulden and D. M. Jackson,
	\textit{The KP hierarchy, branched covers, and triangulations},  Advances in Mathematics,
	{\bf 219}  pp. 932-951, 2008
	
	
\bibitem{GrinevichOrlov} P.G. Grinevich, A.Yu. Orlov, ``Virasoro Action on Riemann Surfaces, Grassmannians,
$\det {\bar\partial}_j$ and Segal-Wilson $\tau$-Function''
 - Problems of Modern Quantum Field Theory, pp.86-106, Springer, Berlin, Heidelberg, 1989;
 P.G. Grinevich, A.Yu. Orlov,
 ``Flag Spaces in KP Theory and Virasoro Action on $\det {\bar \partial}_j$ and Segal-Wilson $\tau$-Function'',
arXiv:9804019



\bibitem{HO-2MM}  J. Harnad and A. Yu. Orlov, ``Fermionic construction of partition functions for two-matrix models
and perturbative Schur function expansions'', J. Phys. A 39, pp. 8783--8809 (2006)


\bibitem{HO-2014} J. Harnad and A. Yu. Orlov, ``Hypergeometric $\tau$-functions, Hurwitz numbers and enumeration of paths'',
Commun. Math. Phys. {\bf 338} (2015) pp. 267--284,
arxiv: math.ph/1407.7800


	\bibitem{HO-2003}  J. Harnad and A. Yu. Orlov, \textit{Scalar product of symmetric functions and matrix integrals},
	Theoretical and mathematical physics 137 (3), pp. 1676-1690  (2003)
	
\bibitem{HO-2006}  J. Harnad and A. Yu. Orlov, \textit{Fermionic construction of partition functions for two matrix models
	and perturbative Schur functions expansions}, J. Phys. A 39, pp. 8783-8809 (2006)
	
	
	\bibitem{HirotaOhta} R.Hirota, Y.Ohta, ``Hierarchies of coupled soliton equations'',
	J.Phys. Soc. Jap. {\bf 60} (1991) pp. 798-809


        \bibitem{JM} M. Jimbo and T. Miwa, ``Solitons and infinite
        dimensional Lie algebras'', Publ. RIMS Kyoto Univ.
        {\bf 19}, pp. 943--1001  (1983)


\bibitem{KvdL}
V. Kac and J. van de Leur, The $n$-component $KP$ hierarchy and representation theory,
Jour. Math. Phys. 44,  3245--3293 (2003).

	
	\bibitem{KvdLbispec} V. Kac and J. van de Leur,
	\textit{The Geometry of Spinors and the Multicomponent BKP and DKP
	Hierarchies}, CRM Proceedings and Lecture Notes {\bf 14}  (1998) pp.
	159-202
	
\bibitem{Kakei-2} S.Kakei, ``Dressing Method and the Coupled KP hierarchy'', arxiv:9909024 (1999)


	\bibitem{Kazakov}
	V. A. Kazakov, M. Staudacher, T. Wynter, \textit{Character Expansion Methods for Matrix Models of Dually Weighted Graphs},
	Commun.Math.Phys. 177 (1996) 451-468;  arXiv:hep-th/9502132 (1995)
	

	\bibitem{Kazakov2} V.A. Kazakov, M. Staudacher and T. Wynter, Ecole Normale preprint LPTENS-95/24,
	hep-th/9506174,  accepted  for  publication  in  Commun. Math. Phys.
	
	\bibitem{Kazakov3} V.A. Kazakov, M. Staudacher and T. Wynter, Ecole Normale preprint LPTENS-95/56,
	CERN preprint CERN-TH/95-352, hep-th/9601069 submitted for publication to Nuclear Physics B.
	
\bibitem{Kazakov-SolvMM} V. A. Kazakov, ``Solvable Matrix Models'', arXiv:hep-th/0003064 (2000)

\bibitem{Kazakov-ZinnJ} V. A. Kazakov and P. Zinn-Justin, ``Two-Matrix model with ABAB interaction'',
        Nucl.Phys. B546 (1999) 647-668

\bibitem{Uspehi-KazarianLando}      M. Kazarian and S. Lando, \textit{Combinatorial solutions to integrable hierarchies},
	Uspekhi Mat. Nauk 70 (2015), no. 3(423), pp. 77-106. English translation: 2015 Russ. Math. Surv. 70, pp. 453-482; arXiv:1512.07172
	
	\bibitem{KazarianLando}   M.  E.  Kazarian  and  S.  K.  Lando,
	\textit{An   algebro-geometric   proof   of Witten's   conjecture},
	J. Amer. Math. Soc. 20:4 (2007), pp. 1079-1089

	
	\bibitem{KZ} M. Kazarian and P. Zograph,  \textit{Virasoro constraints and topological recursion for
	Grothendieck's dessin counting},
	arxiv{1406.5976}

	
\bibitem{KMMM}  S. Kharchev, A. Marshakov, A. Mironov and A. Morozov,
\textit{Generalized Kazakov-Migdal-Kontsevich Model: group theory aspects},
International Journal of Mod Phys A10 (1995) p.2015

\bibitem{Kharchev98} S. Kharchev, ``Kadomtsev-Petviashvili Hierarchy and Generalized Kontsevich Model'',
arXiv:hep-th/9810091 

\bibitem{KMMOZ}  K. Kharchev, A. Marshakov, A. Mironov, A. Orlov, A. Zabrodin, 
``Matrix models among integrable theories:
Forced hierarchies and operator formalism'', Nuclear Physics B 366, 569-601, 1991

\bibitem{Mironych-2-komp}  S. Kharchev, A. Marshakov, A. Mironov, A. Morozov, S. Pakuliak,
Conformal Matrix Models as an Alternative to Conventional Multi-Matrix Models, Nucl.Phys.B404:717-750,1993


\bibitem{KrichMineevZabr} I. Krichever, M. Mineev-Weinstein, P. Wiegmann, A. Zabrodin,
Laplacian Growth and Whitham Equations of Soliton Theory,
Physica D198 (2004) pp. 1-28; arXiv:nlin/0311005

\bibitem{ZL} S. K. Lando, A. K. Zvonkin, ``Graphs on Surfaces and their Applications'', Encyclopaedia of Mathematical Sciences,
 Volume 141, with  appendix by D. Zagier, Springer, N.Y. (2004)

\bibitem{LO-LMP} J.W. van de Leur, A. Yu. Orlov, ``Pfaffian and determinantal Tau functions''
Letters in Mathematical Physics 105 (11) (2015) 1499-1531 

\bibitem{L1} J. W. van de Leur, \textit{Matrix Integrals and Geometry of 	Spinors}, { J. of Nonlinear Math. Phys.} {\bf  8},  pp. 288-311 (2001)

\bibitem{LN} S.Loktev, Natanzon S.M., Klein topological field theories from group representations, SIGMA, 7(2011), paper 070, 15 pp.

\bibitem{Mac} I.G. Macdonald, {Symmetric Functions and Hall Polynomials},
	Clarendon Press, Oxford, (1995).

	
	
	\bibitem{Mehta} M. L. Mehta ``Random Matrices'', 3nd edition (Elsevier, Academic, San Diego CA, 2004)
	
	
\bibitem{Migdal} A.A.Migdal, JETP, 42, 413 (1975).

\bibitem{Mikhailov} A.V. Mikhailov, ``On the Integrability of two-dimensional Generalization of the 
Toda Lattice'', Letters in Journal of Experimental and Theoretical Physic
s, v.30, p. 443-448, 1979;
	A.V.Mikhailov, M. A. Olshanetski, A.M.Perelomov, Two-dimentional generalized Toda lattice, Comm.Math.Phys {\bf 79} (1981) no. 4 473-488
	
	\bibitem{MWZ}
	M.Mineev-Weinstein, P.Wiegmann, A.Zabrodin,
	\textit{Integrable Structure of Interface Dynamics},
	Phys. Rev. Lett. 84 (2000) 5106-5109



\bibitem{MMS} A. Mironov,A., Morozov and G. Semenoff,  Unitary Matrix Integrals in the Framework of
the Generalized Kontsevich Model, Intern J Mod Phys A 11 (1996) 5031-5080


	
\bibitem{MM1} A. D. Mironov, A. Yu. Morozov and S. M. Natanzon, \textit{Complect set of cut-and-join operators in the
	Hurwitz-Kontsevich theory}, Theor. and Math.Phys. 166:1,(2011), pp.1-22; arXiv:0904.4227
	

\bibitem{MM2} A. D. Mironov, A. Yu. Morozov and S. M. Natanzon,
	\textit{A Hurwitz theory avatar of open-closed strings}, The European Physical Journal C73 (2013) 2324

	
\bibitem{MM3} A. D. Mironov, A. Yu. Morozov and S. M. Natanzon,
	\textit{Algebra of differential operators associated with Young diagramms}, J.Geom.and Phys. n.62(2012), pp. 148-155

 
\bibitem{MM4} A. D. Mironov, A. Yu. Morozov and S. M. Natanzon,
 \textit{Integrability of Hurwitz Partition Functions. I. Summary}, J. Phys. A: Math. Theor. 45 (2012) 045209
	
	
\bibitem{MM5} A. D. Mironov, A. Yu. Morozov and S. M. Natanzon,
 \textit{Integrability properties of Hurwitz partition functions. II. Multiplication of cut-and-join operators and WDVV equations}, JHEP 11 (2011) 097

\bibitem{Morozov-MM-review} A.Yu.Morozov, Integrability and Matrix Models, Uspehi Fizicheskih
Nauk, vol 164 (1994) pp.3-62

\bibitem{N90} S. M. Natanzon, ``Klein surfaces'', Russian Math. Surv. 45:6(1990), pp. 53--108

\bibitem{Nat} S. M. Natanzon, ``Extended cohomological field theories and noncommutative Frobenius manifods'', J.Geom. Phys. 51(2004) no.4, pp. 387--403


\bibitem{N1} Natanzon S.M., Cyclic foam topological field theory, J.Geom.Phys. 60(2010), no.6-8, 874-883, arXiv:0712.3557

\bibitem{N2004} S. M. Natanzon, \textit{Moduli of Riemann surfaces, real algebraic curves and their superanalogs},
	Translations of Math. Monograph, AMS, Vol.225 (2004), 160 p.

	
\bibitem{N} S. M. Natanzon, \textit{Simple Hurwitz numbers of a disk}, Funk. Analysis ant its applications, v.44 (2010), n1, pp. 44-58
	
	\bibitem{NO-2014}  S. M. Natanzon and A. Yu. Orlov, \textit{Hurwitz numbers and BKP hierarchy},
	arXiv:1407.832
	
	\bibitem{NO-LMP} S. M. Natanzon and A. Yu. Orlov, \textit{BKP and projective Hurwitz numbers},
	Letters in Mathematical Physics, 107(6), 1065-1109 (2017); arXiv:1501.01283
	
	\bibitem{NatanzonZabrodin}  S. M. Natanzon and A. Zabrodin,
	\textit{Toda hierarchy, Hurwitz numbers and conformal dynamics},
	Int. Math. Res. Notices 2015 (2015) 2082-2110

\bibitem{NovikovManakovZakharov} Theory of Solitons, V.Zakharov, S.Manakov, S.Novikov, L.Pitaevskii, Nauka 1979

	\bibitem{Okounkov-2000} A. Okounkov, \textit{Toda equations for Hurwitz numbers}, { Math. Res. Lett.} {\bf 7}, pp. 447-453 (2000).
	See also arxiv{math-004128}
	

	
	\bibitem{Okounkov-Pand-2006} A. Okounkov and R. Pandharipande,
	\textit{Gromov-Witten theory, Hurwitz theory and completed cycles},
	Annals of Math {\bf 163} p.517 (2006); arxiv.math.AG/0204305
	
\bibitem{Orl1987} A. Yu. Orlov, ``Vertex operators, ${\bar\partial}$-problem, symmetries, variational identities and Hamiltonian formalism
for 2+1 integrable systems'', Nonlinear and Turbulent Processes in Physics, ed. V. Baryakhtar. Singapore: World Scientific, 1988

\bibitem{OS-2000} A. Yu. Orlov and D. Scherbin, \textit{Fermionic representation for basic hypergeometric
functions related to
Schur polynomials}, arXiv preprint nlin/0001001

\bibitem{Rusakov}  Rusakov

\bibitem{OS-PhysicaD}
A. Yu. Orlov and D. Scherbin, ``Multivariate hypergeometric functions  as $\tau$-functions for Toda lattice and KP
equations'', Physica D, vol 152-152 pp 51-65 (2001)

\bibitem{Milne} A. Yu. Orlov and D. Scherbin, ``Milne's hypergeometric functions in terms of free fermions'',
Journal of Physics A: Mathematical and General 34 (11), p.~2295;
S. C. Milne, ``Summation theorems for basic hypergeometric series of Schur function argument'',
in Progress in Approximation Theory, Eds. A. A. Gonchar and E. B. Saff, pp.
51--77, Springer-Verlag, New York (1992)

\bibitem{OS-TMP} A. Yu. Orlov and D. Scherbin,
\textit{Hypergeometric solutions of soliton equations},
Theoretical and Mathematical Physics 128 (1), pp. 906-926 (2001)

\bibitem{O-Acta} A. Yu. Orlov, \textit{Soliton theory, symmetric functions and matrix integrals},
Acta Applicandae Mathematica 86 (1-2), pp. 131-158 (2005); 

\bibitem{O-review2002} A.Yu.Orlov, ``Tau functions and matrix integrals'', arXiv preprint math-ph/0210012;


\bibitem{O-TMP-2006} A.Yu.Orlov, ``Hypergeometric functions as infinite-soliton tau functions'',
Theoretical and Mathematical Physics, 146 (2): 183-206 (2006);
A. Yu. Orlov, ``Hypergeometric tau functions $\tau(t, T, t^*)$ as $\infty$-soliton
tau function in $T$ variables,'' nonlin.SI/0305001 (2003).

\bibitem{O-2004-New} A. Yu. Orlov, \textit{New solvable matrix integrals} ,
{  Intern.  J. Mod. Phys.} {\bf  A 19 } (suppl 02), pp. 276-93  (2004).


\bibitem{OShiota-2004} A. Yu. Orlov and T. Shiota, \textit{Schur function expansion for normal matrix model and associated
discrete matrix models}, Physics Letters A 343 (5), pp. 384-396

\bibitem{OST-I} A. Yu. Orlov, T. Shiota and K. Takasaki, \textit{Pfaffian structures and certain solutions to
BKP hierarchies I. Sums over partitions},  arXiv: math-ph/12014518;

\bibitem{OST-II} A. Yu. Orlov, T. Shiota and K. Takasaki, Pfaffian structures and certain solutions to
  BKP hierarchies II. Multiple integrals, arXiv preprint arXiv:1611.02244

\bibitem{O-2012} A. Yu. Orlov, ``Deformed Ginibre ensembles and integrable systems'',
Physics Letters A 378, pp. 319--328 (2014)

\bibitem{O-TMP-2017} A.Yu.Orlov \textit{Hurwitz numbers and products of random matrices},
Theoretical and Mathematical Physics 193(3) pp 1282-1323 (2017); arxiv:1701.02296;

\bibitem{O-2017} A.Yu.Orlov, ``Links between quantum chaos and counting problems'', arXiv preprint arXiv:1710.10696, 2017

\bibitem{PogrebkovSushko} A. K. Pogrebkov and V. N. Sushko, `` Quantization of the $(sin\psi)_2$ interaction in
terms of fermion variables'', Translated from Teoretieheskaya i Mathematicheskaya Fizika,
Vol. 24, No. 3, pp.425-429, September, 1975. Original article submitted May 15, 1975


\bibitem{Rumanov} I. Rumanov, ``Classical integrability for beta ensembles and general Fokker-Plank equation'',
Journal of Math Phys {\bf 56} 013508 (2015)


\bibitem{Sato} M. Sato and Y. Sato, Soliton equations as dynamical systems on infinite-dimensional
Gr
Grassmann manifold. Nonlinear partial differential equations in applied science
(Tokyo 1982), 259-271
North-Holland Math. Stud., 81, North-Holland, Amsterdam, 1983.

\bibitem{S1} E. Strahov, \textit{Dynamical correlation functions for products of random matrices}, arXiv:1505.02511

\bibitem{S2} E. Strahov, \textit{Differential equations for singular values of products of Ginibre random matrices},
arXiv:1403.6368

\bibitem{Takasaki-Schur} K. Takasaki, Initial value problem for the
Toda lattice hierarchy, {\it Adv. Stud. Pure Math.} {\bf 4}
(1984) 139--163

\bibitem{Takasaki2018} K.Takasaki, ``Toda hierarchies and their applications'',   J. Phys. A: Math. Theor. 51 (2018),
203001 (35pp)

\bibitem{TI} T. Takebe,Representation Theoretical Meaning
of Initial Value Problem for the Toda Lattice Hierarchy I, {\it
LMP\/} {\bf 21} (1991) 77--84

\bibitem{TII} T. Takebe,  Representation Theoretical Meaning
of Initial Value Problem for the Toda Lattice Hierarchy II, {\it
Publ. RIMS, Kyoto Univ.} {\bf 27} (1991), 491--503


\bibitem{VKIII} N.Ya. Vilenkin and A.U. Klimyk,
{\it Representation of Lie Groups and Special Functions. Volume 3:
Classical and Quantum Groups and Special Functions}, Kluwer
Academic Publishers, 1992

\bibitem{UT} K.Ueno and K.Takasaki, \textit{Toda lattice hierarchy},
{\it Adv. Stud. Pure Math.} {\bf  4},  1-95 (1984).

\bibitem{Witten} E.Witten, ``On Quantum Gauge Theories in Two Dimensions'', Com.Math.Phys. {\bf 141}
153-209 (1991)

\bibitem{ZakharovShabat} V. E. Zakharov and A. B. Shabat,
J. Funct. Anal. Appl. 8, 226 (1974), 13, p. 166 (1979)

\bibitem{ZinnJustin} P.  Zinn-Justin, HCIZ integral and 2D Toda lattice hierarchy.
Nuclear Physics B vol 634
Issue 3 (2002) pp 417-432;
P. Zinn-Justin and J.B. Zuber, On some integrals over the U(N) unitary group and their
large limit, math-ph/0209019

		
\bibitem{Zog}  P. Zograf, Enumeration of Grothendieck's dessins and KP hierarchy,
Int. Math. Res. Notices 24, 13533-13544 (2015);  arXiv:1312.2538 (2013).









	


\end{thebibliography}
\end{document}